\documentclass[10pt,journal]{IEEEtran}
\usepackage{amsmath,amsfonts}

\usepackage{amsthm,amssymb}

\usepackage{bm}
\usepackage{algorithm}
\usepackage{algpseudocode}
\usepackage{array}
\usepackage{subfigure}
\usepackage{textcomp}
\usepackage{stfloats}
\usepackage{url}
\usepackage{verbatim}
\usepackage{graphicx}
\usepackage{cite}
\usepackage{color}

\begin{document}

\title{Optimal Transceiver Design in Over-the-Air Federated Distillation}

\author{\normalsize Zihao Hu, \textit{Student Member, IEEE}, Jia Yan, \textit{Member, IEEE}, Ying-Jun Angela Zhang, \textit{Fellow, IEEE}, Jun Zhang, \textit{Fellow, IEEE}, and Khaled B. Letaief, \textit{Fellow, IEEE}\thanks{This work was presented in part at the IEEE Vehicular Technology Conference (VTC), Singapore, 2024 \cite{ref17}. Z. Hu and Y.-J. A. Zhang are with the Department of Information Engineering, The Chinese University of Hong Kong, Hong Kong SAR. (e-mail: hz021@ie.cuhk.edu.hk; yjzhang@ie.cuhk.edu.hk). J. Yan is with the Intelligent Transportation Thrust, The Hong Kong University of Science and Technology (Guangzhou), Guangzhou, China. (e-mail: jasonjiayan@hkust-gz.edu.cn). J. Zhang and K. B. Letaief are with the Department of Electrical and Computer Engineering, The Hong Kong University of Science and Technology, Hong Kong SAR. (e-mail: eejzhang@ust.hk; eekhaled@ust.hk).}}

\maketitle
\begin{abstract}
The rapid proliferation and growth of artificial intelligence (AI) has led to the development of federated learning (FL). FL allows wireless devices (WDs) to cooperatively learn by sharing only local model parameters, without needing to share the entire dataset. However, the emergence of large AI models has made existing FL approaches inefficient, due to the significant communication overhead required. In this paper, we propose a novel over-the-air federated distillation (FD) framework by synergizing the strength of FL and knowledge distillation to avoid the heavy local model transmission. Instead of sharing the model parameters, only the WDs' model outputs, referred to as knowledge, are shared and aggregated over-the-air by exploiting the superposition property of the multiple-access channel. We shall study the transceiver design in over-the-air FD, aiming to maximize the learning convergence rate while meeting the power constraints of the transceivers. The main challenge lies in the intractability of the learning performance analysis, as well as the non-convex nature and the optimization spanning the whole FD training period. To tackle this problem, we first derive an analytical expression of the convergence rate in over-the-air FD. Then, the closed-form optimal solutions of the WDs' transmit power and the estimator for over-the-air aggregation are obtained given the receiver combining strategy. Accordingly, we put forth an efficient approach to find the optimal receiver beamforming vector via semidefinite relaxation. We further prove that there is no optimality gap between the original and relaxed problem for the receiver beamforming design. Numerical results will show that the proposed over-the-air FD approach achieves a significant reduction in communication overhead, with only a minor compromise in testing accuracy compared to conventional FL benchmarks.
\end{abstract}
\begin{IEEEkeywords}
Federated distillation, over-the-air computation, edge intelligence, beamforming, semidefinite relaxation.
\end{IEEEkeywords}
\section{Introduction}
The rapid progress of artificial intelligence (AI) has enabled a proliferation of intelligent edge services, such as autonomous driving \cite{ref15} and virtual reality \cite{ref16}. These AI services rely on machine learning models trained over distributed data across wireless devices (WDs) \cite{ref27}. However, the extensive data exchange required for AI model training poses significant challenges for resource-constrained wireless communication systems \cite{ref26}. To address this problem, federated learning (FL) has been proposed as an efficient paradigm for distributed AI model training. With the coordination of a parameter server (PS), WDs cooperatively train a global AI model by sharing their local updates, e.g., weights of the local models \cite{ref0}. With the raw data kept at the WDs, FL significantly reduces the communication overhead while preserving the data privacy.\par
Even though FL only requires communicating local model updates, the massive number of participating WDs limits the communication efficiency and scalability of FL. Recognizing that the PS is primarily interested in the aggregated result of local updates from the WDs, over-the-air computation has been proposed as a way to enhance the communication efficiency of FL \cite{ref1}. Specifically, the WDs can simultaneously transmit the analog modulated local updates using the same radio-frequency resources, leveraging the inherent superposition property of the multiple-access channel to attain the desired model aggregation \cite{ref28}. Apart from the single-antenna setting \cite{ref1, ref22}, recent studies have harnessed the spatial diversity in multi-antenna over-the-air FL systems to further improve the learning performance. For the single-input multiple-output (SIMO) setting, the authors in \cite{ref3} jointly designed the receiver beamforming, as well as the user scheduling strategy in over-the-air FL to minimize the distortion of aggregated signals. In \cite{ref23}, a joint device selection and receiver beamforming approach was proposed to maximize the number of participants under the model aggregation error constraint. In addition, the co-design of the uplink and downlink communication schemes was studied for SIMO over-the-air FL in \cite{ref24}. In \cite{ref2}, over-the-air FL under the multiple-input multiple-output (MIMO) channel was considered, where the receiver beamforming and learning rate are jointly optimized to improve the learning performance.\par
Aggregating local updates over-the-air, however, still incurs a significant uplink communication overhead due to the widespread adoption of large AI models, e.g., large language models (LLMs) with hundreds of millions of parameters for the discourse analysis and real-time conversations \cite{ref29}. \textcolor{black}{To mitigate the huge communication overhead in over-the-air FL, the authors in \cite{ref22} proposed to sparsify the gradient before the uplink transmission, followed by which an approximate message passing algorithm is leveraged for the gradient reconstruction at the PS. Model pruning technique is adopted in \cite{ref33}, which concurrently alleviates the computation load at WDs, and enhances the communication efficiency through model size reduction. Nevertheless, due to the ever-increasing size of learning models, the communication efficiency achieved by these light-weight over-the-air FL approaches becomes insignificant, while the learning performance is severely degraded.} The stringent latency requirements and scarce radio resources call for a cooperative learning paradigm that bypasses the transmission of massive model parameters. Federated distillation (FD) becomes a favorable candidate, with the synergy of FL and knowledge distillation \cite{ref4}. The key idea of knowledge distillation is to incorporate the model output, referred to as ``knowledge", into the training loss function as the regularizer \cite{ref5}. In FD, only the model outputs of WDs, a.k.a., soft predictions, are transmitted and aggregated. Instead of sharing the model updates per training round, each WD in FD extracts the knowledge of the other devices' datasets from the aggregated soft predictions. Accordingly, the dimension of the uplink transmission signal of each WD will only depend on the number of sample classes in the training datasets, regardless of the AI model size. Towards communication-efficient FD, the authors in \cite{ref6} proposed compressed federated distillation by quantizing and delta-coding the soft predictions before the uplink transmission through orthogonal channels. A selective knowledge sharing mechanism was proposed in \cite{ref14} to enhance the performance of aggregation, which also reduces the communication overhead. To accommodate communication imperfections in the FD framework with massive MIMO, the authors in \cite{ref25} proposed a group split mechanism, where the received soft predictions at the PS are divided into two groups based on the corresponding channel conditions. Moreover, the authors in \cite{ref7} leveraged over-the-air computation for the aggregation of soft predictions from the WDs to further reduce the communication overhead in the FD systems.\par
Over-the-air FD opens up potential benefits for improving the communication efficiency when training large-size AI models cooperatively over wireless networks. \textcolor{black}{For the synchronization of the teacher and student model outputs, the previous literature on FD typically assumes a public dataset at the PS, which not only fails to preserve the data privacy of WDs, but also requires sharing the model outputs as many as the public dataset size. Further, the existing works focus on the FD design over error-free communication link. Notably, the aggregation of local model outputs suffers from inevitable perturbations due to the wireless channel fading and communication noise in practical FD systems. To mitigate the corresponding performance degradation, it is necessary to study the optimal transceiver design when deploying the FD framework over wireless networks. In this paper, we consider the FD systems without the public dataset by periodically exchanging the per-label local-averaged soft prediction in the presence of wireless channel fading and communication noises. Over-the-air computation is further leveraged to improve the communication efficiency of FD over resource-constrained large-scale wireless networks. In addition, we theoretically analyze the impact of imperfect wireless channels and optimize the transceiver design accordingly to improve the performance of the considered over-the-air FD system. To our best knowledge, this is the first attempt to study the optimal transceiver design for over-the-air FD, which is not a trivial extension of the existing works on over-the-air FL.}\par
\textcolor{black}{The fundamental differences between the FD and FL frameworks render the existing over-the-air results inapplicable. Specifically, local model parameters or the corresponding gradients are aggregated per round in FL, whereas each WD in the considered FD system only uploads the local-averaged model outputs per label with the model kept locally. Accordingly, the FL system restricts the local model structure to be the same, while local model can be diverse in the considered FD system. In addition, the contribution of each WD’s dataset is controlled by the weight during the model averaging in FL. Instead, each WD in FD leverages a distillation regularizer during the local training to capture the knowledge of the other WDs’ local datasets, wherein the distillation weight balances the local and distilled information in the local loss function.}\par
\textcolor{black}{In this regard, the existing convergence analysis in over-the-air FL literature, especially characterizing the impact of the wireless channel fading and communication noises on the learning performance, cannot be directly applied in the FD context. The unique challenges of this work are threefold. First, apart from upper bounding the model update each round using convexity and smoothness assumptions on the training loss function as the conventional FL analysis does, the convergence analysis in FD further requires to bound the additional distillation regularizer, which is generally non-convex. Moreover, the wireless channel fading and communication noises perturb the model output knowledge presented in the regularizer, further exacerbating the analysis. Second, model outputs aggregated over-the-air may deviate from the target global-averaged knowledge due to various transmit power and channel fading across WDs. It is challenging to analyze the impact of such knowledge misalignment error on the FD performance. Notice that the degradation of the learning performance caused by such inherent signal misalignment in the conventional over-the-air FL is dominated by the straggler with the worst channel condition and power capacity. In contrast, the corresponding learning performance gap in over-the-air FD is affected by multiple stragglers, one for each data class. With local-averaged model soft predictions aggregated per label class, the diverse per-class local dataset sizes and various per-label local knowledge statistics among WDs result in distinct knowledge misalignment across different classes. Lastly, after tackling the challenges in over-the-air FD convergence analysis, solving the corresponding optimal transceiver design problem is also difficult due to the non-convex nature of the problem as well as the strong coupling between over-the-air power control and receive beamforming.}\par
To tackle the aforementioned challenges, this paper makes the first attempt to study the transceiver optimization for learning performance maximization in over-the-air FD. The problem is formulated by maximizing the convergence rate subject to the power constraints of the transceivers. The contributions of this paper are summarized as follows:
\begin{itemize}
    \item We consider an over-the-air FD system, where WDs simultaneously transmit soft predictions using the same radio resources for over-the-air aggregation at the multi-antenna PS. Each WD updates its model parameters through gradient descent with the loss function involving the globally aggregated soft prediction as the regularizer.
    \item We derive the closed-form expression of the convergence rate for over-the-air FD with a diminishing step size under mild assumptions. Accordingly, we formulate a transceiver optimization problem over the transmit power of the WDs, estimator for over-the-air aggregation, and the receiver combining scheme, by maximizing the convergence rate of the considered FD system.
    \item To tackle the challenging non-convex transceiver design problem, we derive the closed-form solutions of the optimal transmit equalization factors and post-processing scalars, given the receiver beamforming strategy.
    \item We design an efficient approach to obtain the optimal receiver combining based on the semidefinite relaxation. We further prove that there is no optimality gap between the original and relaxed optimization problems over the receiver beamforming vector. 
\end{itemize}
Simulation results will demonstrate that our proposed over-the-air FD approach exhibits superior learning performance while achieving significant communication overhead reduction compared with conventional FL alternatives.\par
The rest of this paper is organized as follows. In Section II, we describe the over-the-air FD model with soft prediction transmission. Then, the communication design problem is formulated in Section III based on the FD convergence analysis. In Section IV, we propose an efficient algorithm to obtain the optimal transceiver design per round. In Section V, we present extensive numerical results to evaluate the proposed method. Finally, this paper concludes in Section VI.
\section{System Model}
\subsection{FD System}
\begin{figure}
\centering
\includegraphics[width=\linewidth]{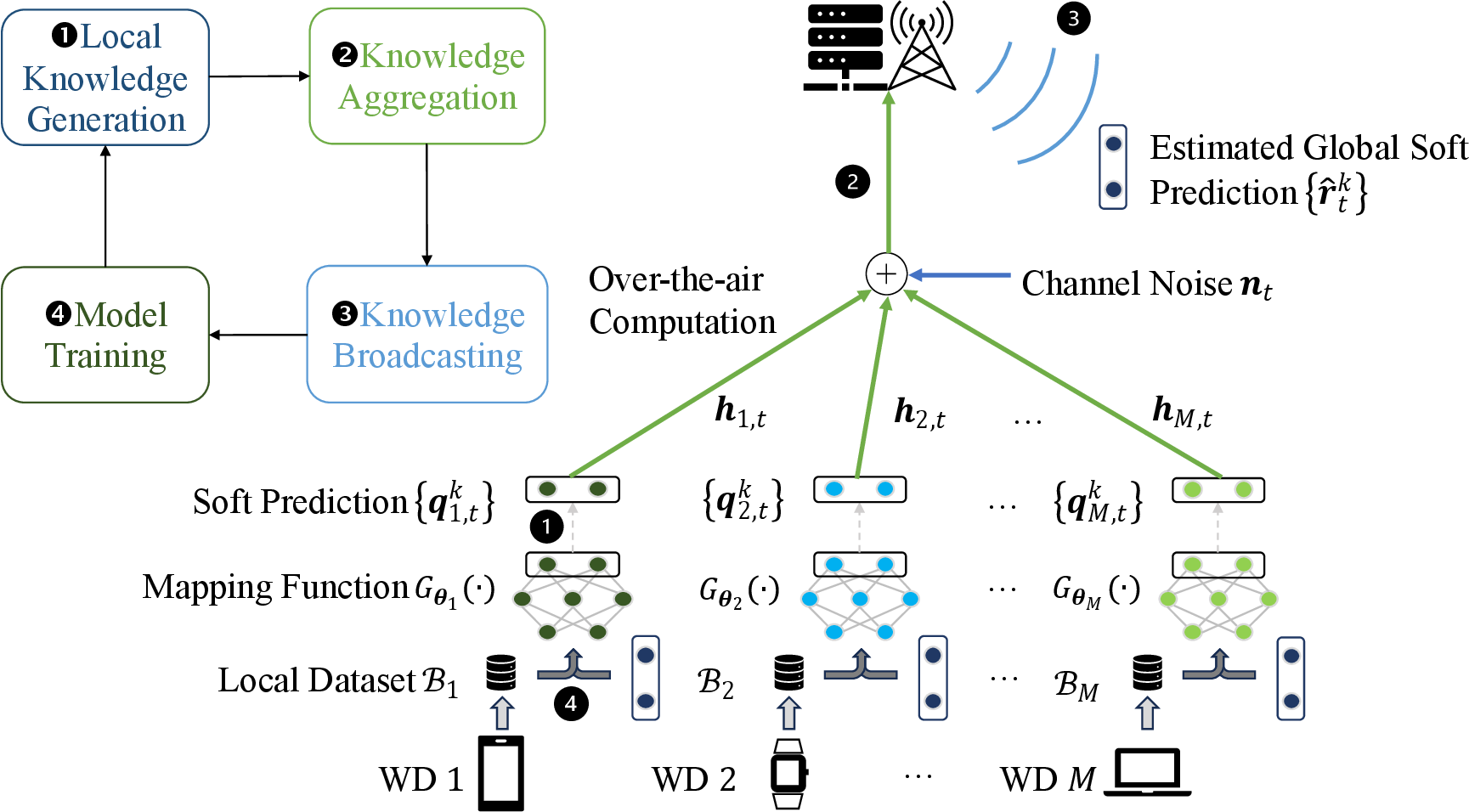}
\caption{\textcolor{black}{The proposed over-the-air FD framework.}}
\end{figure}
Consider an FD system comprising an \(N\)-antenna PS and \(M\) single-antenna WDs, collected in the index set \(\mathcal{M}=\{1,2,\cdots,M\}\). WD \(i\) possesses a local dataset \(\mathcal{B}_i\) consisting of \(B_i\) training samples. There are a total of \(K\) data sample classes in each local training dataset, e.g., 10 classes from digit 0 to 9 presented in the MNIST database for hand-writing digit classification, or clustering labels into \(K\) groups for a regression task. We partition the local dataset \(\mathcal{B}_i\) into \(K\) disjoint subsets, namely \(\mathcal{B}_i^1,\cdots,\mathcal{B}_i^k,\cdots,\mathcal{B}_i^K\), where the subset \(\mathcal{B}_i^k\) contains all the training samples belonging to class \(k\) in \(\mathcal{B}_i\). Additionally, \(B_i^k=\left|\mathcal{B}_i^k\right|\) representing the cardinality of the local subset \(\mathcal{B}_i^k\). Each WD \(i\) aims to learn a global model parameterized by \(\bm{\theta}\) without sharing its local model parameter \(\bm{\theta}_i\) or the corresponding gradient as the traditional FL does. To achieve this goal, each WD \(i\) leverages online co-distillation \cite{ref5} under the coordination of the PS and targets at minimizing its own empirical local loss function, i.e.,
\begin{equation}
\begin{aligned}
    \min_{\bm{\theta}\in\mathbb{R}^D}F_i\left(\bm{\theta};\left\{\mathbf{q}^{v_{i,b}}\right\}_{v_{i,b}=1}^K\right)&=\frac{1}{B_i}\sum_{b=1}^{B_i}f\left(\bm{\theta};\mathbf{u}_{i,b},v_{i,b}\right)\\
    &\quad+\gamma\left\|G_{\bm{\theta}}\left(\mathbf{u}_{i,b}\right)-\mathbf{q}^{v_{i,b}}\right\|^2_2,
\end{aligned}
\end{equation}
where \(D\) is the dimension of the model parameter. Likewise, \(\left(\mathbf{u}_{i,b},v_{i,b}\right)\) is the \(b\)-th training sample in \(\mathcal{B}_i\), where \(\mathbf{u}_{i,b}\) denotes the feature vector and \(v_{i,b}\) represents the output label. \(G_{\bm{\theta}}(\mathbf{u}_{i,b})\in \mathbb{R}^K\) is the model output, i.e., soft prediction, with respect to the input \(\mathbf{u}_{i,b}\). As will be specified later, \(\mathbf{q}^{v_{i,b}}\) is the global-averaged soft prediction for training samples belonging to class \(v_{i,b}\) from all WDs, referred to as the distilled knowledge. \(f\left(\cdot\right)\) is a general training loss function while the distillation loss function is chosen as the \(l_2\) norm loss function, which quantifies the distance between the model output \(G_{\bm{\theta}}\left(\cdot\right)\) and distilled knowledge. \(\gamma\) is the weight of the distillation loss.\par
To address Problem (1), WD \(i\) employs a gradient descent algorithm to update the model parameter \(\bm{\theta}_{i,t}\) iteratively with \(T\) training rounds. Specifically, as depicted in Fig. 1, the \(t\)-th training round, \(1\leq t\leq T\), consists of the following steps:
\begin{itemize}
    \item \textit{Local knowledge generation:} WD \(i\) computes the local-averaged soft prediction vector for training samples in each class \(k\) based on the current local model parameter \(\bm{\theta}_{i,t}\), i.e,
\begin{equation}
    \mathbf{q}_{i,t}^k=\frac{1}{B_i^k}\sum_{b\in\mathcal{B}_i^k}G_{\bm{\theta}_{i,t}}\left(\mathbf{u}_{i,b}\right).
\end{equation}
\(\mathbf{q}_{i,t}^k\) is also referred to as the local knowledge of class \(k\) for WD \(i\) at training round \(t\).
    \item \textit{Knowledge aggregation:} WDs transmit the local knowledge \(\left\{\left\{\mathbf{q}_{i,t}^k\right\}_{k=1}^K\right\}_{i=1}^M\) to the PS through the wireless channels. For global knowledge aggregation of each class \(k\), the PS tends to compute the global-averaged soft prediction vector
    \begin{equation}
        \mathbf{q}^k_t=\sum_{i=1}^M\frac{B_i^k}{B^k}\mathbf{q}_{i,t}^k,
    \end{equation}
    where \(B^k=\sum_{i=1}^MB_i^k\) is the total number of training samples that belongs to class \(k\) in all WDs. Nevertheless, due to the channel fading and communication noises, the PS can only obtain an estimate of \(\mathbf{q}^k_t\), denoted by \(\mathbf{\hat{r}}^k_t\), which is used for the model update at each WD.
    \item \textit{Knowledge broadcasting:} The PS broadcasts the estimated global knowledge \(\left\{\mathbf{\hat{r}}_t^k\right\}_{k=1}^K\) to all WDs through an error-free link \cite{ref2,ref3}. Notice that the error-free downlink broadcast transmission is an acceptable approximation for most practical scenarios since the PS with a stable power supply generally has adequate communication capability to perform the error correction.
    \item \textit{Model training:} WD \(i\) updates its model parameters via gradient descent, i.e.,
    \begin{equation}
    \bm{\theta}_{i,t+1}=\bm{\theta}_{i,t}-\eta_t\nabla F_i\left(\bm{\theta}_{i,t};\left\{\mathbf{\hat{r}}_t^k\right\}_{k=1}^K\right),
\end{equation}
where \(\eta_t\) is the learning rate at training round \(t\).
\end{itemize}
\subsection{Transmit Signal Design}
At the beginning of the training round \(t\), \(1\leq t\leq T\), each WD \(i\) generates the local knowledge by (2) and computes the statistics of the local-averaged soft prediction vectors (i.e., the means and variances) for normalization:
\begin{equation}\small
    \overline{q}_{i,t}^k=\frac{1}{K}\sum_{d=1}^K q_{i,t}^k[d],\quad\hat{q}_{i,t}^{k^2}=\frac{1}{K}\sum_{d=1}^K \left(q_{i,t}^k[d]-\overline{q}_{i,t}^k\right)^2,\;\forall k,
\end{equation}
where \(q_{i,t}^k[d]\) is the \(d\)-th entry of \(\mathbf{q}_{i,t}^k\). These scalar quantities \(\left\{\overline{q}_{i,t}^k,\hat{q}_{i,t}^k\right\}\) are uploaded to the PS for post-processing. Accordingly, WD \(i\) normalizes each local-averaged soft prediction by
\begin{equation}
    \mathbf{x}^k_{i,t} = \frac{\mathbf{q}_{i,t}^k-\overline{q}_{i,t}^k\mathbf{1}}{\hat{q}_{i,t}^k},\quad\forall k,
\end{equation}
where \(\mathbf{1}\in\mathbb{R}^K\) is the \(K\)-dimensional all-one column vector. Then, the transmit signal of WD \(i\) is given by
\begin{equation}\small
\hat{\mathbf{x}}_{i,t}=\left[\left(P^1_{i,t}\mathbf{x}^1_{i,t}\right)^\top, \left(P^2_{i,t}\mathbf{x}^2_{i,t}\right)^\top,\cdots, \left(P^K_{i,t}\mathbf{x}^K_{i,t}\right)^\top\right]^\top\in\mathbb{C}^{K^2},
\end{equation}
where \(P^k_{i,t}\in\mathbb{C}\) is the transmit equalization factor for the normalized local-averaged soft prediction vector \(\mathbf{x}_{i,t}^k\) at training round \(t\). The normalization step in (6) guarantees that
\begin{equation}\small
    \mathbb{E}\left[\left|\hat{x}_{i,t}[d]\right|^2\right]=\left|P^k_{i,t}\right|^2,\;\forall d\in\left[\left(k-1\right)K+1, kK\right],i,k,t.
\end{equation}
Here, \(\hat{x}_{i,t}[d]\) denotes the \(d\)-th entry of \(\hat{\mathbf{x}}_{i,t}\). Accordingly, \(\left|P_{i,t}^k\right|^2\) controls the transmit power of WD \(i\) for the local-averaged soft prediction of class \(k\). In this paper, we consider the peak transmit power constraint of each WD as
\begin{equation}
    \left|P_{i,t}^k\right|^2\leq P_i, \quad\forall i,k,t,
\end{equation}
where \(P_i\) is the peak transmit power of WD \(i\).
\vspace{-0.4cm}
\subsection{Soft Prediction Aggregation via Over-the-Air Computation}
With over-the-air computation \cite{ref1}, all WDs synchronously transmit \(\left\{\hat{\mathbf{x}}_{i,t}\right\}_{i=1}^M\) entry by entry using the same radio-frequency resources, over a total of \(K^2\) time slots\footnote{Notice that the number of classes \(K\) is several orders of magnitude smaller than the dimension of DNN model parameters in typical image classification tasks. This demonstrates significant savings in communication resources for our proposed approach compared to the traditional over-the-air FL.}. As a result, the transmitted signals naturally superpose in the multiple-access channel. In this paper, we assume a block fading channel model, where the channel coefficients remain constant throughout an FD training round and may vary across different rounds. Additionally, we assume perfect channel state information (CSI) at the PS per training round. The CSI can be estimated via uplink channel estimation at the training phase of each fading state \cite{ref20, ref21}. The channel coefficient vector between the single-antenna WD \(i\) and the \(N\)-antenna PS at training round \(t\) is denoted as \(\mathbf{h}_{i,t}\in\mathbb{C}^N\). With over-the-air computation, at the \(d\)-th time slot, the PS receives
\begin{equation}
    \mathbf{y}_t[d]=\sum_{i=1}^M\mathbf{h}_{i,t}\hat{x}_{i,t}[d]+\mathbf{n}_t[d],
\end{equation}
where \(\mathbf{n}_t[d]\sim\mathcal{CN}\left(\mathbf{0},\sigma_n^2\cdot \mathbf{I}\right)\) is the additive white Gaussian noise (AWGN) vector at the \(d\)-th time slot. Each entry has mean \(0\) and variance \(\sigma_n^2\). To exploit the space diversity, the PS combines the received signals from \(N\) antennas by a normalized receiver beamforming vector \(\mathbf{w}_t\in\mathbb{C}^N\) with \(\left\|\mathbf{w}_t\right\|^2_2=1,\forall t\),
\begin{equation}
    \hat{y}_t[d]=\sum_{i=1}^M\mathbf{w}_t^H\mathbf{h}_{i,t}\hat{x}_{i,t}[d]+\mathbf{w}_t^H\mathbf{n}_t[d].
\end{equation}
The received signal corresponding to the aggregated soft prediction of class \(k\) is given by
\begin{equation}\small
    \mathbf{r}^k_t=\left[\hat{y}_t\left[\left(k-1\right)K+1\right],\cdots,\hat{y}_t\left[kK\right]\right]^\top,\quad\forall k\in\left[1,K\right].
\end{equation}
Then, the PS estimates the desired global-averaged soft prediction vector \(\mathbf{q}^k_t\) for each class \(k\) by a linear estimator, i.e.,
\begin{equation}
    \begin{aligned}
            \mathbf{\hat{r}}^k_t&=\frac{\mathbf{r}^k_t}{\lambda^k_t}+\sum_{i=1}^Ma^k_{i,t}\overline{q}_{i,t}^k\mathbf{1}=\sum_{i=1}^Ma^k_{i,t}\overline{q}_{i,t}^k\mathbf{1}\\
            &\quad+\sum_{i=1}^M\frac{\mathbf{w}_t^H\mathbf{h}_{i,t}P_{i,t}^k}{\lambda^k_t\hat{q}_{i,t}^k}\left(\mathbf{q}_{i,t}^k-\overline{q}_{i,t}^k\mathbf{1}\right)+\frac{\hat{\mathbf{n}}^k_t}{\lambda^k_t},\quad \forall k,
    \end{aligned}
\end{equation}
where \(\hat{\mathbf{n}}^k_t=\left[\mathbf{w}_t^H\mathbf{n}_t\left[\left(k-1\right)K+1\right],\cdots,\mathbf{w}_t^H\mathbf{n}_t\left[kK\right]\right]^\top,\forall k\). \(\lambda^k_t\) and \(a^k_{i,t}\) are post-processing normalization scalars of class \(k\) for WD \(i\) at training round \(t\). The PS then broadcasts estimated signals \(\left\{\mathbf{\hat{r}}^k_t,\forall k\right\}\) to all WDs for the model update in (4).
\section{Convergence Analysis and Problem Formulation}
In this section, we first analyze the convergence performance of the considered over-the-air FD system. Aiming to maximize the convergence performance, we then formulate the transceiver design problem.
\vspace{-0.4cm}
\subsection{Convergence Analysis}
We characterize the convergence performance of the FD system by the asymptotical expected gradient norm of the empirical local loss functions. To begin with, we have the following assumptions.\par
\textit{\textbf{Assumption 1:}} The gradients of the local loss functions \(F_i(\cdot)\) are \(L_1\)-Lipschitz continuous, i.e.,
\begin{equation}\small
\left\|\nabla F_i(\bm{\theta}_{i,t_1})-\nabla F_i(\bm{\theta}_{i,t_2})\right\|_2\leq L_1\left\|\bm{\theta}_{i,t_1}-\bm{\theta}_{i,t_2}\right\|_2,\;\forall i,t_1, t_2.
\end{equation}
\par
\textit{\textbf{Assumption 2:}} The learned model function mapping \(G_{\bm{\theta}}(\cdot)\) at each WD is \(L_2\)-Lipschitz continuous, i.e.,
 \begin{equation}\small
     \left\|G_{\bm{\theta}_{i,t_1}}(\cdot)-G_{\bm{\theta}_{i,t_2}}(\cdot)\right\|_2\leq L_2\left\|\bm{\theta}_{i,t_1}-\bm{\theta}_{i,t_2}\right\|_2,\;\forall i,t_1, t_2.
 \end{equation}\par
\textit{\textbf{Assumption 3:}} The gradient norm of the loss function is uniformly bounded by a scalar \(S\), i.e.,
\begin{equation}
    \left\|\nabla F_i\left(\bm{\theta}_{i,t}\right)\right\|_2\leq S,\quad \forall i,t.
\end{equation}\par
Assumption 1 is adopted in \cite{ref11,ref12} and holds for widely-used loss functions in practical learning tasks, such as mean squared error or logistic regression. Assumption 2 and 3 are adopted in \cite{ref11} and hold for many kinds of neural networks, e.g., multi-layer perceptron or convolutional neural network.\par
Under Assumptions 1-3, we derive an upper bound of the expected gradient norm for each WD in the following theorem.\par
\textit{\textbf{Theorem 1:}} Suppose that Assumptions 1-3 hold and the empirical local loss function for WD \(i\) is upper bounded by \(f_{i,max}\), i.e., \(F_i\left(\bm{\theta}\right)\leq f_{i,max},\forall \bm{\theta}\). For any \(\gamma>0\) and initial learning rate \(\eta_0>0\), if \(\eta_t=\frac{\eta_0}{\sqrt{t}}\leq\frac{1}{L_1}\), we have\footnote{\textcolor{black}{Notice that the results in Theorem 1 can be applied to both independent and identically distributed (IID) and non-IID dataset settings.}}\\
\begin{equation}
\begin{aligned}
    \mathbb{E}\left[\left\|\nabla F_i\left(\hat{\bm{\theta}}_{i,T}\right)\right\|^2_2\right]&\leq\frac{3f_{i,max}}{\eta_0\sqrt{T}}+\sum_{t=0}^{T-1}\frac{6\gamma\eta_0L_2\left(L_1\eta_t+1\right)}{\eta_t}\\
    &\quad\times\frac{\left\|\nabla F_i(\bm{\theta}_{i,t})\right\|_2\Phi_{1,i,t}}{T^{\frac{3}{2}}}+8\gamma L_2S\\
    &\quad+\sum_{t=0}^{T-1}6\eta_0\gamma^2L_2^2L_1\left(\frac{\Phi_{1,i,t}^2+\Phi_{2,i,t}^2}{T^{\frac{3}{2}}}\right),
\end{aligned}
\end{equation}
where
\begin{equation}
    \begin{aligned}
        \Phi_{1,i,t} &= \sum_{k=1}^K\frac{B_i^k}{B_i}\left\|\sum_{i=1}^M\left(\frac{\mathbf{w}_t^H\mathbf{h}_{i,t}P_{i,t}^k}{\lambda_t^k\hat{q}_{i,t}^k}-\frac{B_i^k}{B^k}\right)\mathbf{q}_{i,t}^k\right.\\
        &\quad\qquad\qquad\left.+\sum_{i=1}^M\left(a^k_{i,t}-\frac{\mathbf{w}_t^H\mathbf{h}_{i,t}P_{i,t}^k}{\lambda_t^k\hat{q}_{i,t}^k}\right)\overline{q}_{i,t}^k\mathbf{1}\right\|_2,\\
        \Phi^2_{2,i,t}&=\mathbb{E}\left[\sum_{k=1}^K\frac{B_i^k}{B_i}\left\|\frac{\hat{\mathbf{n}}^k_t}{\lambda^k_t}\right\|^2_2\right].
    \end{aligned}
\end{equation}
Here, we denote \(\Phi_{1,i,t}\) and \(\Phi_{2,i,t}\) as the abbreviations of \(\Phi_{1,i,t}\left(\mathbf{w}_t,\left\{P_{i,t}^k,\lambda_t^k,a^k_{i,t},\forall i,k\right\}\right)\) and \(\Phi_{2,i,t}\left(\mathbf{w}_t,\left\{\lambda_t^k,\forall k\right\}\right)\), respectively. Moreover, \(\hat{\bm{\theta}}_{i,T}\) is randomly chosen from \(\left\{\bm{\theta}_{i,t}\right\}_{t=0}^{T-1}\) with probability \(P\left(\hat{\bm{\theta}}_{i,T}=\bm{\theta}_{i,t}\right)=\frac{1/\eta_t}{\sum_{t=0}^{T-1}1/\eta_t}\).\par
\begin{proof}
    See Appendix A.
\end{proof}
We observe that the error function \(\Phi_{1,i,t}\left(\mathbf{w}_t,\left\{P_{i,t}^k,\lambda_t^k,a^k_{i,t},\forall i,k\right\}\right)\) is due to the possible signal misalignment caused by the channel fading. \(\Phi_{2,i,t}\left(\mathbf{w}_t,\left\{\lambda_t^k,\forall k\right\}\right)\) arises from the perturbation of the Gaussian channel noise\footnote{\textcolor{black}{The convergence result in Theorem 1 and the resultant optimal transceiver design in Section IV can be extended to the over-the-air FD system with a heavy-tailed noise \cite{ref35}. Due to the page limit, we omit the details here.}}.
\subsection{Problem Formulation}
With Theorem 1, we formulate the transceiver design problem by minimizing the upper bound of the gradient norm in (17) of all WDs. For brevity, we define \(A_1=6\gamma\eta_0L_2\) and \(A_2=6\eta_0\gamma^2L_2^2L_1\). Our goal is to optimize the transceiver design \(\mathcal{P}_t=\left\{\mathbf{w}_t,P_{i,t}^k,\lambda_t^k,a^k_{i,t},\forall i,k\right\}\) at each round subject to the peak transmit power constraint in (9), i.e.,
\begin{equation}
    \begin{aligned}
        \mbox{(P1)}\quad& \underset{\left\{\mathcal{P}_t\right\}}{\text{min}}
& & \sum_{i=1}^M\sum_{t=0}^{T-1}A_1\frac{\left(L_1\eta_t+1\right)\left\|\nabla F_i(\bm{\theta}_{i,t})\right\|_2\Phi_{1,i,t}}{\eta_tT^{\frac{3}{2}}}\\
 &&   &\quad+\sum_{i=1}^M\sum_{t=0}^{T-1}A_2\frac{\Phi_{1,i,t}^2+\Phi_{2,i,t}^2}{T^{\frac{3}{2}}}\\
 & \text{s.t.} & & \left|P_{i,t}^k\right|^2\leq P_i,\quad \forall i,k,t,\\
 &&& \left\|\mathbf{w}_t\right\|^2_2=1,\quad\forall t.
    \end{aligned}
\end{equation}
Problem (P1) is challenging to solve due to its non-convex nature and the need to optimize over \(T\) training rounds. In the following section, by decomposing Problem (P1) into a per-slot optimization problem, we first obtain the optimal transmit equalization factors \(\left\{P_{i,t}^k, \forall i,k\right\}\) and post-processing scalars \(\left\{\lambda_t^k, a^k_{i,t}, \forall i,k\right\}\) per training round given the receiver beamforming vector \(\mathbf{w}_t\). Then, the optimization problem over \(\mathbf{w}_t\) is efficiently solved through the semidefinite relaxation (SDR) technique \cite{ref8}.
\section{Optimal Transceiver Design}
In this section, we propose an efficient algorithm to obtain the closed-form optimal solutions of the WDs’ transmit power and the estimator for over-the-air aggregation at each training round, together with the optimal receiver beamforming vector via SDR. Then, we analyze the complexity of the proposed algorithm.
\subsection{Optimal Solutions}
Suppose that the receiver beamforming vector \(\mathbf{w}_t\) is given at training round \(t\). We first obtain the optimal transmit equalization factors \(\left\{P_{i,t}^k, \forall i,k\right\}\) and post-processing scalars \(\left\{\lambda_t^k, a^k_{i,t}, \forall i,k\right\}\) in the following proposition.\par
\textit{\textbf{Proposition 1:}} Per training round \(t\), given the receiver beamforming vector \(\mathbf{w}_t\), the optimal transmit equalization factors \(\left\{P_{i,t}^k, \forall i,k\right\}\) are
\begin{equation}
    P_{i,t}^{k^\ast} = \frac{B^k_i\lambda_t^k\hat{q}_{i,t}^k\left(\mathbf{w}_t^H\mathbf{h}_{i,t}\right)^H}{B^k\left|\mathbf{w}_t^H\mathbf{h}_{i,t}\right|^2},\quad\forall i,t,k.
\end{equation}
Moreover, the optimal post-processing scalars \(\left\{\lambda_t^k, a^k_{i,t}, \forall i,k\right\}\) are
\begin{equation}
    \begin{aligned}
        \lambda_t^{k^\ast} &= \min_{i\in\mathcal{M}} \frac{B^k\left|\mathbf{w}_t^H\mathbf{h}_{i,t}\right|\sqrt{P_i}}{B^k_i\hat{q}_{i,t}^k},\quad\forall t,k,\\
        a_{i,t}^{k^\ast} &= \frac{B_i^k}{B^k},\quad\forall i,t,k.
    \end{aligned}
\end{equation}\par
\begin{proof}
    See Appendix B.
\end{proof}
From Proposition 1, we observe that if WD \(i\) holds more data samples of class \(k\), then the associated transmit power is larger. On the other hand, a better channel condition leads to a lower transmit power of WD \(i\).\par
With the optimal transmit equalization factors and post-processing scalars obtained in Proposition 1, the problem with respect to the receiver beamforming vector \(\mathbf{w}_t\) becomes
\begin{equation}
    \begin{aligned}
        \mbox{(P2)}\quad& \underset{\left\{\mathbf{w}_t\right\}}{\text{min}}
& & \sum_{i=1}^M\frac{A_2K\sigma_n^2\sum_{k=1}^KB_i^k\max_{j\in\mathcal{M}} \frac{B^k_j\hat{q}_{j,t}^k}{B^k\left|\mathbf{w}_t^H\mathbf{h}_{j,t}\right|\sqrt{P_j}}}{B_i\sqrt{T}}\\
 & \text{s.t.} & & \left\|\mathbf{w}_t\right\|^2_2=1,\quad\forall t.
    \end{aligned}
\end{equation}
Equivalently, Problem (P2) can be rewritten as
\begin{equation}
    \begin{aligned}
        \mbox{(P3)}\quad& \underset{\left\{\mathbf{w}_t\right\}}{\text{min}}
& & \frac{A_2K\sigma_n^2}{\sqrt{T}}\sum_{i=1}^M\sum_{k=1}^K\frac{B_i^k}{B_iB^k}\max_{j\in\mathcal{M}}-\frac{\left|\mathbf{w}_t^H\mathbf{h}_{j,t}\right|\sqrt{P_j}}{B^k_j\hat{q}_{j,t}^k}\\
 & \text{s.t.} & & \left\|\mathbf{w}_t\right\|^2_2=1,\quad\forall t.
    \end{aligned}
\end{equation}
By defining slack variables
\begin{equation}
    e^k= \max_{j\in\mathcal{M}}-\left|\mathbf{w}_t^H\hat{\mathbf{h}}^k_{j,t}\right|^2,\quad\forall k,
\end{equation}
where \(\hat{\mathbf{h}}^k_{j,t}=\frac{\sqrt{P_j}}{B^k_j\hat{q}_{j,t}^k}\mathbf{h}_{j,t}\), we can transform Problem (P3) into a semidefinite programming problem, i.e.,\\
\begin{subequations}
    \begin{align}
        \mbox{(P4)}\quad& \underset{\left\{\mathbf{W}_t\in\mathcal{H}^N,\left\{e^k\right\}_{k=1}^K\right\}}{\text{min}}
& & \sum_{k=1}^K\frac{e^k}{B^k}\sum_{i=1}^M\frac{B_i^k}{B_i}\tag{25}\\
 & \text{s.t.} & & \mbox{Tr}\left(\mathbf{W}_t\right)=1,\quad\forall t,\tag{25a}\\
 &&&\mathbf{W}_t\succeq\mathbf{0},\quad\forall t,\tag{25b}\\
 &&&\mbox{rank}\left(\mathbf{W}_t\right)=1,\quad\forall t,\tag{25c}\\
 &&&e^1+\mbox{Tr}\left(\mathbf{W}_t\mathbf{H}_{j,t}^1\right)\geq 0,\quad\forall j,t,\tag{25d.1}\\
 &&&\cdots\notag\\
 &&& e^K+\mbox{Tr}\left(\mathbf{W}_t\mathbf{H}_{j,t}^K\right)\geq 0,\quad\forall j,t.\tag{25d.\(K\)}
    \end{align}
\end{subequations}
Here, \(\mathbf{W}_t=\mathbf{w}_t\mathbf{w}_t^H\), \(\mathbf{H}^k_{j,t}=\hat{\mathbf{h}}^k_{j,t}\left(\hat{\mathbf{h}}^k_{j,t}\right)^H\). \(\mathcal{H}^N\) is the set of hermitian \(N\times N\) matrices. By relaxing the non-convex constraint \(\mbox{rank}\left(\mathbf{W}_t\right)=1,\forall t\), Problem (P4) is a standard positive semidefinite programming problem, which can be solved efficiently via off-the-shelf optimization tools, e.g., CVX \cite{ref9}\footnote{\textcolor{black}{We see that the optimal transceiver design in (20), (21), and Problem (P4) are independent of the learning statistical parameters such as \(L_1\) and \(L_2\). Therefore, the proposed optimal transceiver design algorithm in over-the-air FD does not need the measurement of these parameters in practice.}}. In the following proposition, we show that there is no optimality gap between the original and relaxed problems in (P4).\par
\textit{\textbf{Proposition 2:}} The optimal solution \(\mathbf{W}_t^\ast\) for the relaxed Problem (P4), without the non-convex constraint (25c), has a rank-one structure.\par
\begin{proof}
    See Appendix C.
\end{proof}
Consequently, the recovered optimal beamforming vector in (P2) is given by
\begin{equation}
    \mathbf{w}^\ast_t=\frac{\sqrt{\alpha}\mathbf{z}}{\left\|\sqrt{\alpha}\mathbf{z}\right\|_2}=\mathbf{z},
\end{equation}
where \(\alpha=1\) is the only non-zero eigenvalue of \(\mathbf{W}^\ast_t\) according to Proposition 2, while \(\mathbf{z}\) is the corresponding eigenvector. The proposed method to solve Problem (P1) is summarized in Algorithm 1, which has the polynomial time complexity\footnote{According to the simulation results, the average run time per round of Algorithm 1 is 0.12s, which is much smaller than a single fading block length \cite{ref13}. This demonstrates that the proposed low-complexity algorithm can be adapted to the time-varying FD system in an online manner.}.
\begin{algorithm}
\caption{The proposed algorithm for Problem (P1)}\label{IA}
\begin{algorithmic}[1]
\For{every training round \(t\) in \(T\)}
\State Solve Problem (P4) without the rank-one constraint to obtain \(\mathbf{W}^\ast_t\);
\State Obtain \(\mathbf{w}^\ast_t\) by (26);
\State Compute \(\left\{P_{i,t}^{k^\ast}, \forall i,k\right\}\) and \(\left\{\lambda_t^{k^\ast}, a^{k^\ast}_{i,t}, \forall i,k\right\}\) according to Proposition 1.
\EndFor
\end{algorithmic}
\end{algorithm}
\subsection{Complexity Analysis}
For Algorithm 1, the computational complexity of solving the relaxed Problem (P4) in the worst case is \(O\left(N^3\right)\) through the interior point method. Then, the beamforming vector recovery step involving the eigendecomposition has the computational complexity \(O\left(N^{2.376}\right)\). According to Proposition 1, the computational complexity of calculating the optimal transceiver design is \(O\left(M\right)\) per training round. Hence, the overall computational complexity of obtaining the optimal transceivers per FD round in Algorithm 1 is \(O\left(N^3+M\right)\). Notice that the PS can execute Algorithm 1 to obtain the optimal transceiver designing variables \(\left\{P_{i,t}^k,\lambda_t^k,a_{i,t}^k,\mathbf{w}_t,\forall i,k\right\}\) per training round \(t\). Then, the PS sends \(\left\{P_{i,t}^{k^\ast},\forall k\right\}\) to each WD \(i\) via a reliable link for proper over-the-air knowledge uploading. This schedule avoids acquisition of CSI at the transmitter (CSIT) every training round to reduce the corresponding communication overhead.
\section{Numerical Results}
In this section, we present extensive simulations to evaluate and compare the proposed over-the-air FD method with the current state-of-the-art approaches.
\subsection{Simulation Setup}
Consider one PS equipped with 5 antennas, and a total number of \(M=50\) WDs in the FD system. We assume that the WDs are located in a shadowed urban area and communicate with the PS through the cellular radio. The distance \(d_i\) between the WD \(i\) and the PS is uniformly distributed from \(100\) meters to \(500\) meters. Each WD has a peak transmit power \(P_i=1\) mW, while the variance of the channel noise is \(\sigma_n^2=10^{-8}\). The channel coefficient is modelled by the small scale fading multiplied by the square root of the path loss. The small scale fading coefficients across different WDs and receive antennas of the PS are IID, following the standard Gaussian distribution. The free-space path loss is given by \(G_{PS}G_{D}\left(\frac{3\times 10^8}{4\pi f_c d_i}\right)^{PL}\), where \(G_{PS}=G_{D}=0\) dB are the antenna gains at the PS and WDs, respectively. \(f_c=915\) MHz is the carrier frequency while we choose \(PL=4\) to be the path loss exponent due to the shadowed urban environment \cite{ref30}.\par
We simulate an image classification task \textcolor{black}{on ResNet-18 over the FMNIST and CIFAR-10 dataset. Specifically, the considered ResNet-18 model comprises of 8 residual blocks, a global average pooling layer, a \(512\times 10\) fully connected layer and finally a softmax layer, with a total of \(D=11178378\) parameters.} The FMNIST dataset consists of 60000 \(28\times28\) grayscale training images in \(K=10\) classes and 10000 testing images. \textcolor{black}{The CIFAR-10 dataset consists of 60000 \(32\times32\) color images in \(K=10\) classes. There are 50000 training images and 10000 testing images.} We set the learning rate decay factor as \(1/\sqrt{t}\), with the initial learning rate \(\eta_0=0.01\). In addition, considering heterogeneous data distribution commonly encountered in FL, we further devise the following IID and non-IID dataset settings \textcolor{black}{for the CIFAR-10 dataset}.\par
\begin{itemize}
    \item \textbf{IID setting}: The training samples are independently and identically distributed among WDs with equal dataset sizes, i.e., \textcolor{black}{\(B_i=\frac{50000}{M}=1000, \forall i\).}
    \item \textbf{Non-IID setting}: \textcolor{black}{A common non-IID dataset partitioning method in the literature is adopted, namely the Dirichlet distribution for quantifying the label distribution skew \cite{ref31}. We choose the parameter \(\psi=0.5\) to simulate the typical non-IID scenario.}
\end{itemize}
\begin{figure}
    \centering
    \includegraphics[width=0.7\linewidth]{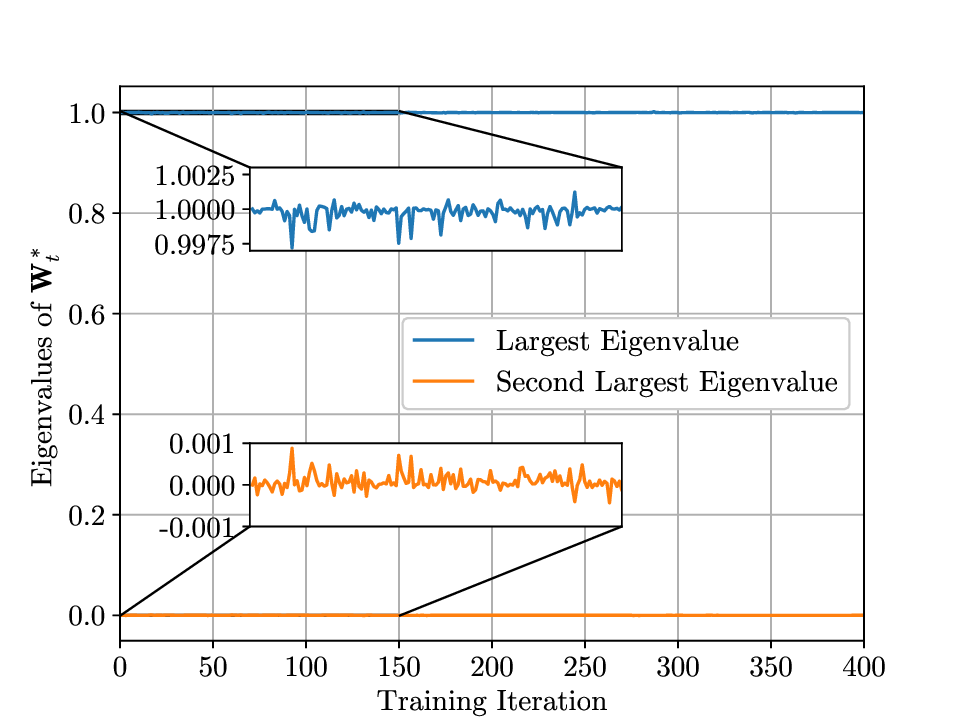}
    \caption{The most significant two eigenvalues of the solution \(\mathbf{W}_t^\ast\) obtained in each round.}
\end{figure}
\vspace{-0.3cm}
\subsection{Verification of Proposition 2}
We first conduct simulations to verify the result in Proposition 2. Specifically, 5 independent experiments are implemented by training ResNet-18 over FMNIST for \(T=400\) rounds. According to Algorithm 1, after solving the relaxed Problem (P4) per round, we obtain \(\mathbf{W}^\ast_t\) and perform the eigendecomposition accordingly. In Fig. 2, we plot the first two largest eigenvalues of \(\mathbf{W}_t^\ast\) per training round. We observe that the most significant eigenvalue of \(\mathbf{W}_t^\ast\) oscillates around the value 1, while the second largest eigenvalue of \(\mathbf{W}_t^\ast\) is in the vicinity of 0. This indicates that \(\mathbf{W}^\ast_t\) obtained from the relaxed Problem (P4) satisfies the rank-one condition, demonstrating no optimality gap between the original and the relaxed Problem (P4). These also coincide with our results in Proposition 2.
\subsection{Effect of Distillation Weight}
\begin{figure}
    \centering
    \includegraphics[width=0.7\linewidth]{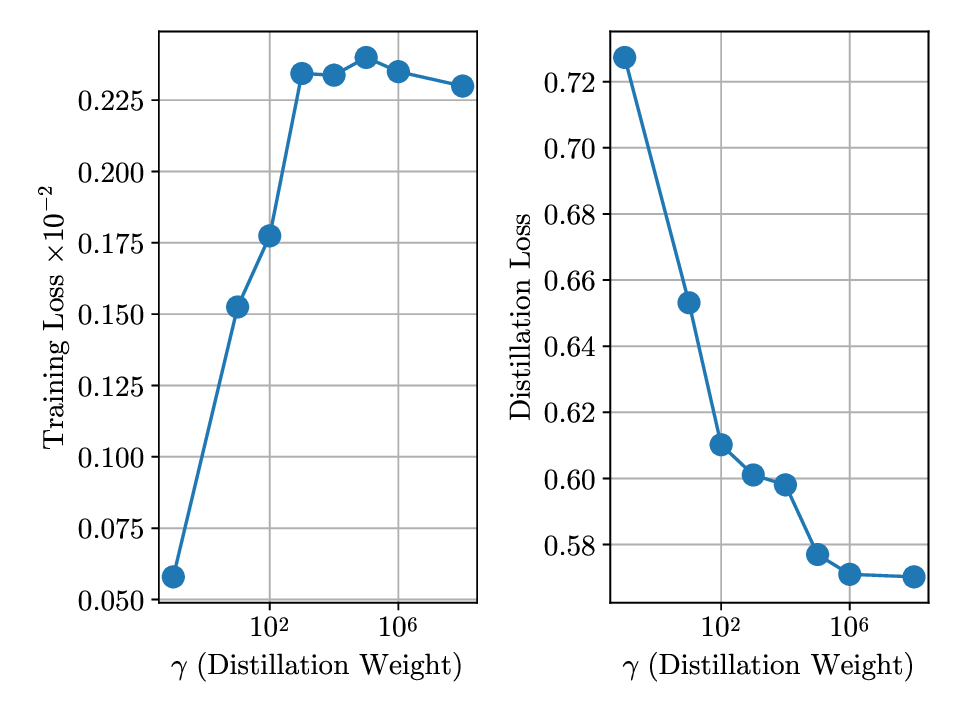}
    \caption{Average training loss and distillation loss versus the distillation weight.}
\end{figure}
\begin{figure}
    \centering
    \subfigure[FMNIST]{\begin{minipage}[b]{0.49\linewidth}
    \centering
            \includegraphics[width=\linewidth]{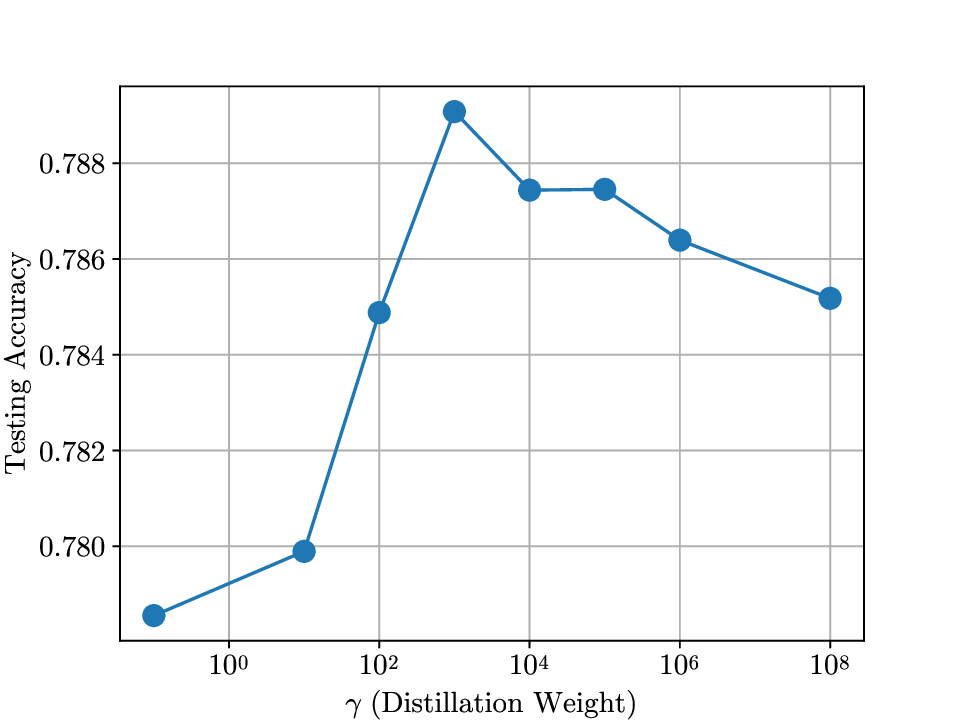}
        \end{minipage}}
        \subfigure[\textcolor{black}{CIFAR-10}]{\begin{minipage}[b]{0.49\linewidth}
            \includegraphics[width=\linewidth]{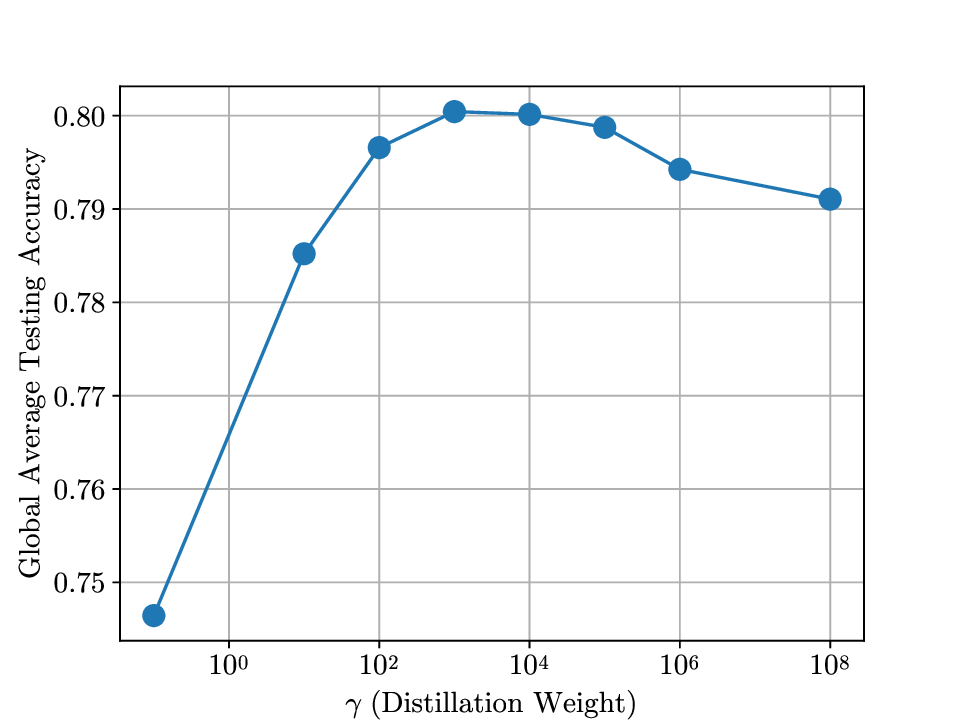}
        \end{minipage}}
    \caption{Average testing accuracy versus the distillation weight.}
\end{figure}
We then study the impact of the distillation weight \(\gamma\) on the learning performance. Notice that $\gamma$ in the local loss function of each WD balances the general training loss (\(f(\cdot)\) in (1)), and the distillation loss. In Fig. 3, we plot the final general training loss and distillation loss under the ResNet18-FMNIST task after \(T=400\) training rounds, as a function of the distillation weight \(\gamma\). It is observed that as the distillation weight increases, the general training loss rises while the distillation loss decreases. This indicates that a larger distillation weight shifts each WD's focus from the individual dataset to the other WDs' knowledge during the training process.\par
Furthermore, Fig. 4 depicts the final testing accuracy as $\gamma$ varies under the ResNet18-FMNIST \textcolor{black}{and CIFAR-10 task after 400 rounds, respectively}. We observe that the testing accuracy initially improves as \(\gamma\) increases. It is because the model parameters at WDs can avoid getting stuck in the local optimum by putting more weight on the distillation loss to learn from the other WDs' local datasets. Then, further increasing \(\gamma\) (e.g., larger than \(1000.0\) in Fig. 4) will lead to a performance drop regarding the final testing accuracy, failing to fully exploit its own local dataset.
\vspace{-0.3cm}
\subsection{Effect of the Number of Antennas}
\begin{figure}
    \centering
    \subfigure[\textcolor{black}{Convergence Gap}]{\begin{minipage}[b]{0.49\linewidth}
    \centering
            \includegraphics[width=\linewidth]{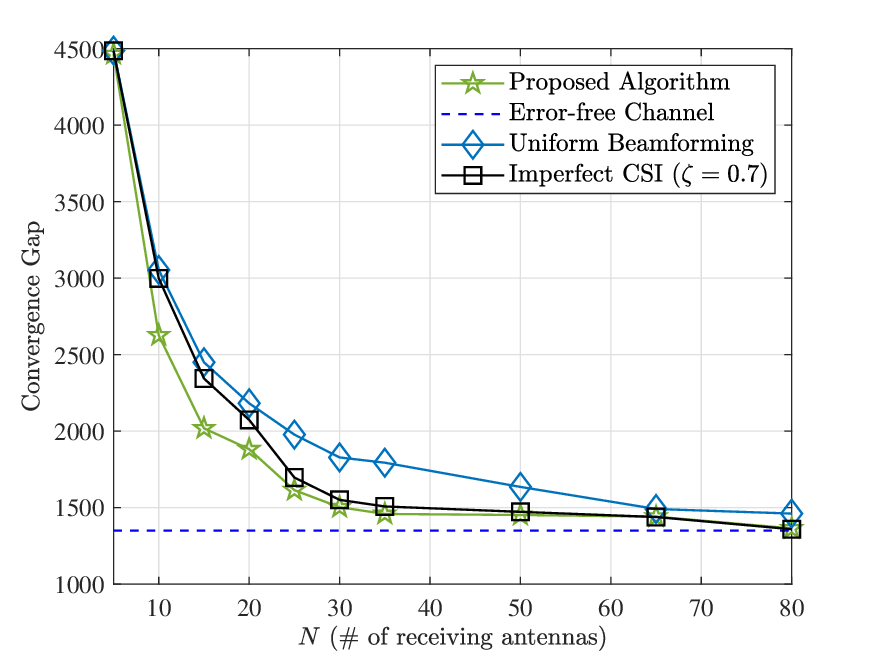}
        \end{minipage}}
        \subfigure[\textcolor{black}{Testing Accuracy}]{\begin{minipage}[b]{0.49\linewidth}
            \includegraphics[width=\linewidth]{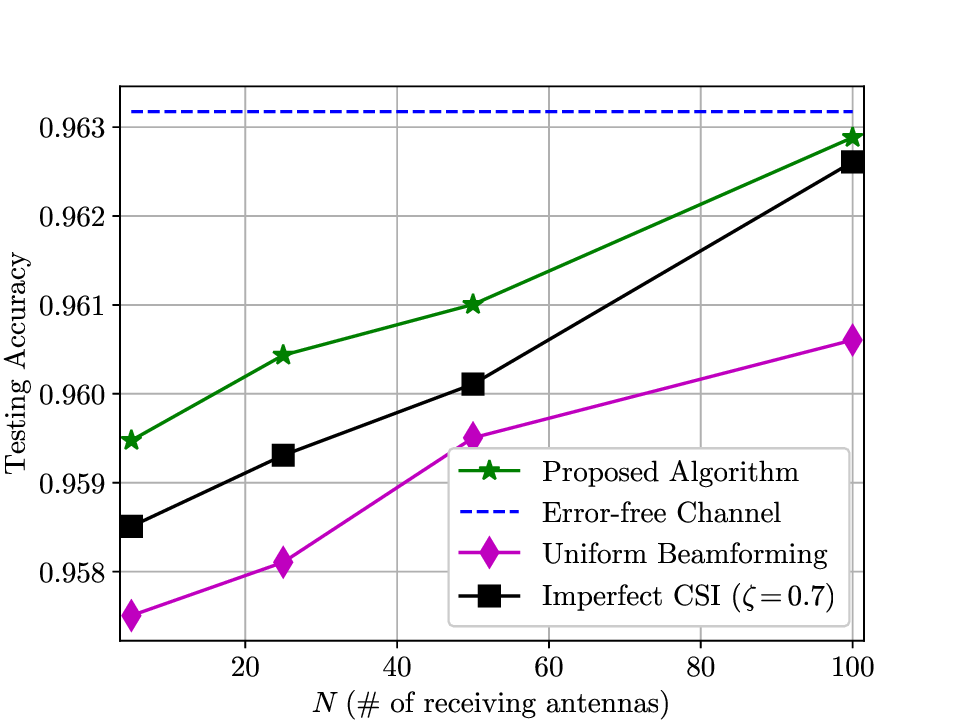}
        \end{minipage}}
    \caption{\textcolor{black}{Analytical convergence gap and average global testing accuracy versus the number of antennas at the PS.}}
\end{figure}
Here we conduct experiments to examine the impact of the number of receiving antennas on the learning performance. In addition to the FedSGD algorithm on the error-free channel as a benchmark, we compare the proposed over-the-air FD approach with the uniform beamforming strategy, where the PS leverages a uniform receiver beamforming vector for combining and WDs upload the local averaged soft predictions with peak transmit power. \textcolor{black}{Moreover, we investigate the robustness of the proposed algorithm against the CSI error on the system optimization. Instead of the perfect CSI estimation, the estimated channel coefficient \(\hat{\mathbf{h}}_{i,t}\) for executing Algorithm 1 is subject to additive errors, which is characterized by
\begin{equation}
    \hat{\mathbf{h}}_{i,t}=\sqrt{\zeta}\mathbf{h}_{i,t}+\sqrt{1-\zeta}\widetilde{\mathbf{n}}_{i,t},\quad\forall i,t.
\end{equation}
Here \(\widetilde{\mathbf{n}}_{i,t}\sim\mathcal{CN}\left(\mathbf{0},\mathbf{I}\right)\) is the independent channel estimation error. \(\zeta\in[0,1]\) stands for the degree of the CSI error.} Fig. 5 (a) presents the analytical convergence gap for all approaches, with respect to the number of antennas at the PS. As the number of antennas at the PS increases, we observe a diminished convergence gap in both the proposed and uniform beamforming approaches. This improvement stems from the fact that a larger antenna array at the PS allows for leveraging more signals to counteract the effect of the channel fading and noise, leading to enhanced learning performance. \textcolor{black}{Meanwhile, the performance gap between the proposed approaches with perfect and imperfect CSI estimation diminishes with the increasing number of receiving antennas, showcasing the robustness of the proposed algorithm with a larger antenna array to combat against the CSI error.}\par
In Fig. 5 (b), we plot the global average testing accuracy of all approaches in the IID setting from 5 independent experiments, with respect to the number of receiving antennas. Similar performance improvement can be observed with the increasing receiving antennas. Notably, the proposed method outperforms the uniform beamforming strategy, asymptotically approaching the performance achieved on the error-free channel. This observation underscores the benefit of the proposed method in effectively harnessing the spatial diversity.
\subsection{Performance Comparison}
\textcolor{black}{We consider the representative benchmarks for performance comparison in the following.
\begin{itemize}
\item \textbf{Error-free FD}: Each WD performs local model training for \(E=5\) iterations by stochastic gradient descent, before uploading their per-label local-averaged model outputs to the PS through error-free channels for global aggregation.
        \item \textbf{Error-free FedAvg}: Each WD performs local model training for \(E=5\) iterations by stochastic gradient descent, before uploading local gradients to the PS through error-free links for global aggregation.
        \item (Over-the-air FD with) \textbf{Uniform Beamforming}: After single local epoch by stochastic gradient descent, each WD uses its respective peak transmit power to upload model outputs to the PS for global aggregation. The PS uses a uniform beamforming vector to receive transmitted signals from WDs.
        \item (FD with) \textbf{Orthogonal Channels}: Each WD uploads the model outputs through orthogonal analog channels after \(E=5\) local iterations by stochastic gradient descent.
        \item (Over-the-air FD with) \textbf{MSE Minimization} \cite{ref23}: Each WD performs local model training for \(E=5\) iterations by stochastic gradient descent. For each communication round, the transmit power for the model outputs of each WD and the beamforming vector at the PS are obtained by minimizing the corresponding MSE.
        \item (Over-the-air FL with) \textbf{90\%-Sparse Rate} \cite{ref22}: Each WD performs local model training for \(E=5\) iterations by stochastic gradient descent. Then, each WD sparsifies 90\% of its local gradient and sends the result to the PS for global aggregation. The PS estimates the transmitted model updates by approximate message passing (AMP) algorithm.
        \item (Over-the-air FL with) \textbf{80\%-P} (pruning) \textbf{and QAT} (quantization-aware training) \cite{ref33}: Before the commence of the FL process, each WD prunes 80\% of its respective local model parameters by magnitude-based pruning. During the FL process, each WD performs local model training with QAT for \(E=5\) iterations by stochastic gradient descent.
    \end{itemize}
\begin{figure}
    \centering
    \subfigure[\textcolor{black}{Average global testing accuracy versus the overall time consumption.}]{\begin{minipage}[b]{\linewidth}
    \centering
            \includegraphics[width=0.7\linewidth]{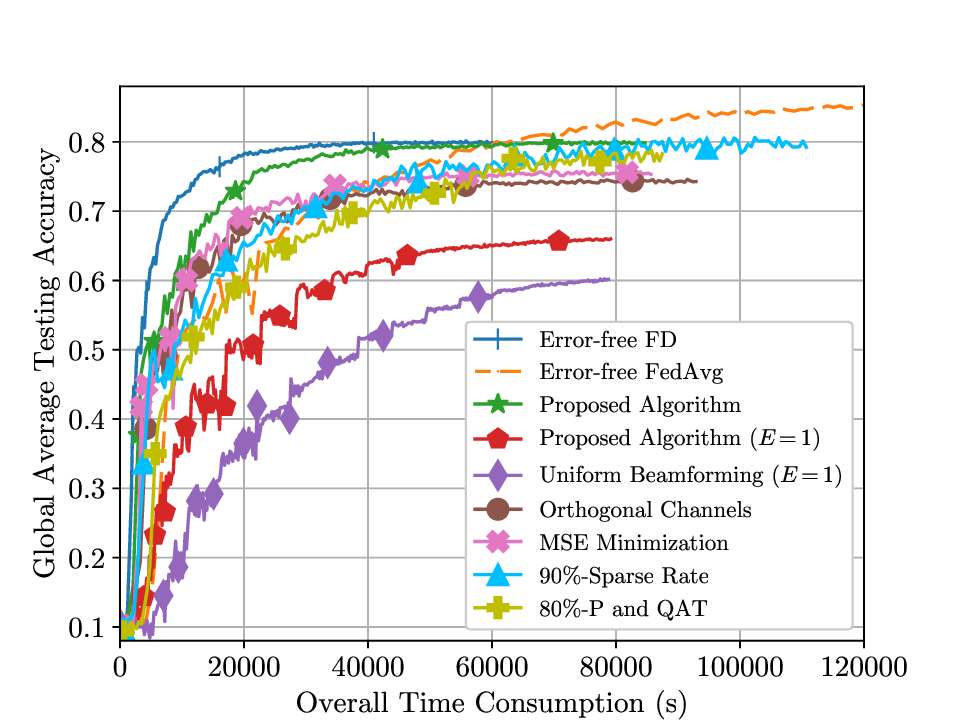}
        \end{minipage}}\\
        \vspace{-0.3cm}
        \subfigure[\textcolor{black}{Average global testing accuracy versus the uplink communication time.}]{\begin{minipage}[b]{\linewidth}
        \centering
            \includegraphics[width=0.7\linewidth]{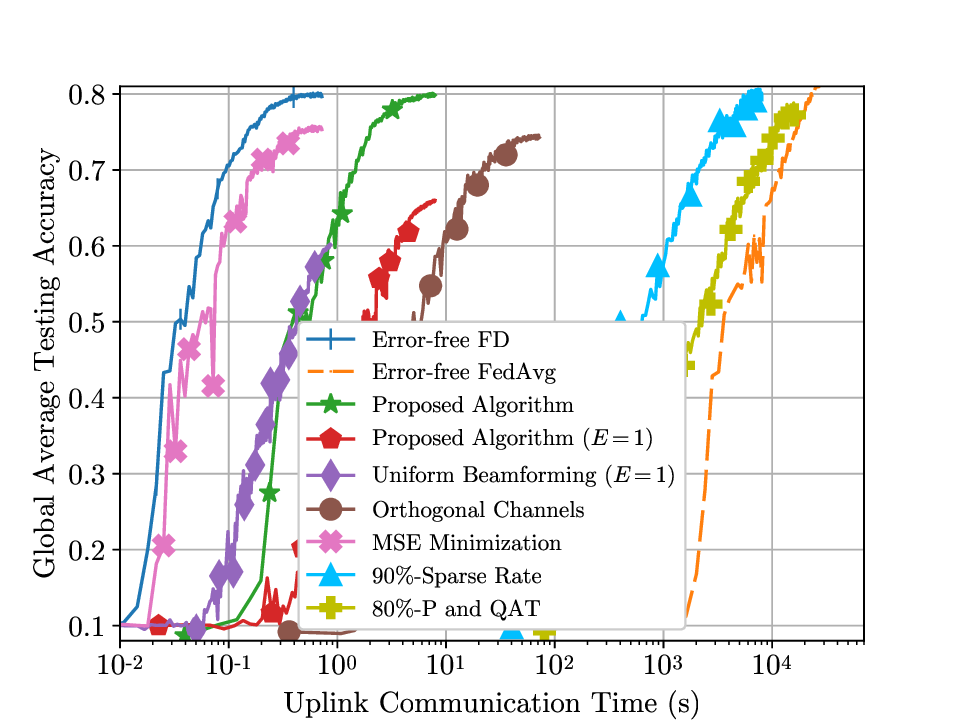}
        \end{minipage}}
    \caption{\textcolor{black}{Average global testing accuracy of ResNet-18 over CIFAR-10 in the IID setting.}}
\end{figure}
Regarding the uplink communication time, the uploading delay for soft predictions as well as the information exchange for executing Algorithm 1 are considered. The uplink communication time is calculated based on real-world communication protocol, i.e., each scalar in the transmitted signal occupies 3.6 \(\mu\)s according to 802.11ac standard \cite{ref10}. The overall time consumption comprises the computation time at both WDs and the PS, together with the uplink communication time.} The performance is averaged over 5 independent simulations.\par
\begin{figure}
    \centering
    \subfigure[\textcolor{black}{Average global testing accuracy versus the overall time consumption.}]{\begin{minipage}[b]{\linewidth}
    \centering
            \includegraphics[width=0.7\linewidth]{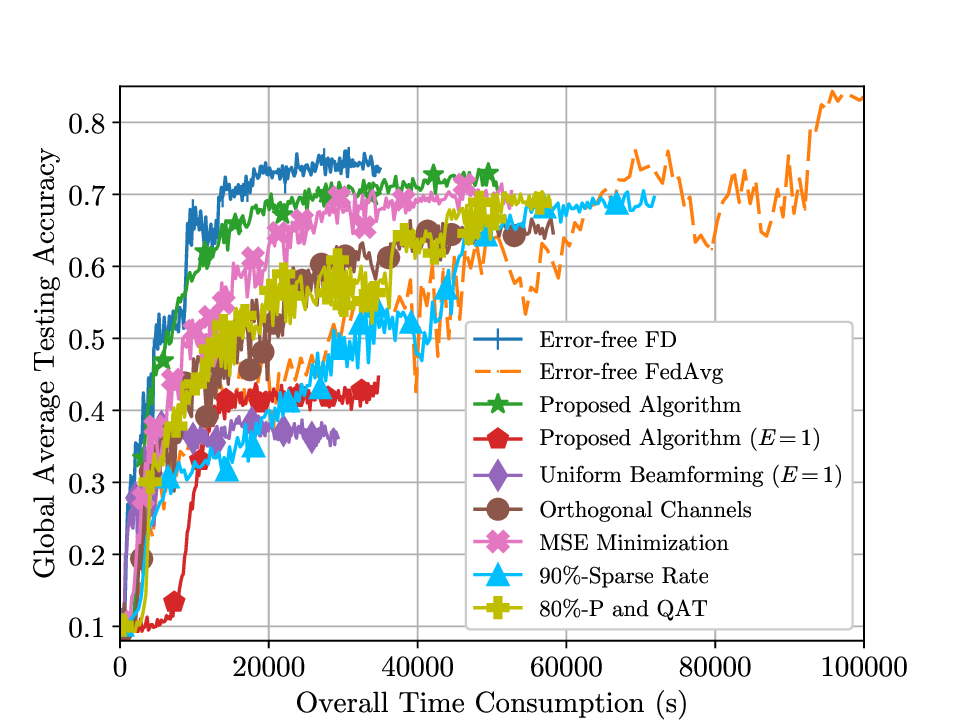}
        \end{minipage}}\\
        \vspace{-0.3cm}
        \subfigure[\textcolor{black}{Average global testing accuracy versus the uplink communication time.}]{\begin{minipage}[b]{\linewidth}
        \centering
            \includegraphics[width=0.7\linewidth]{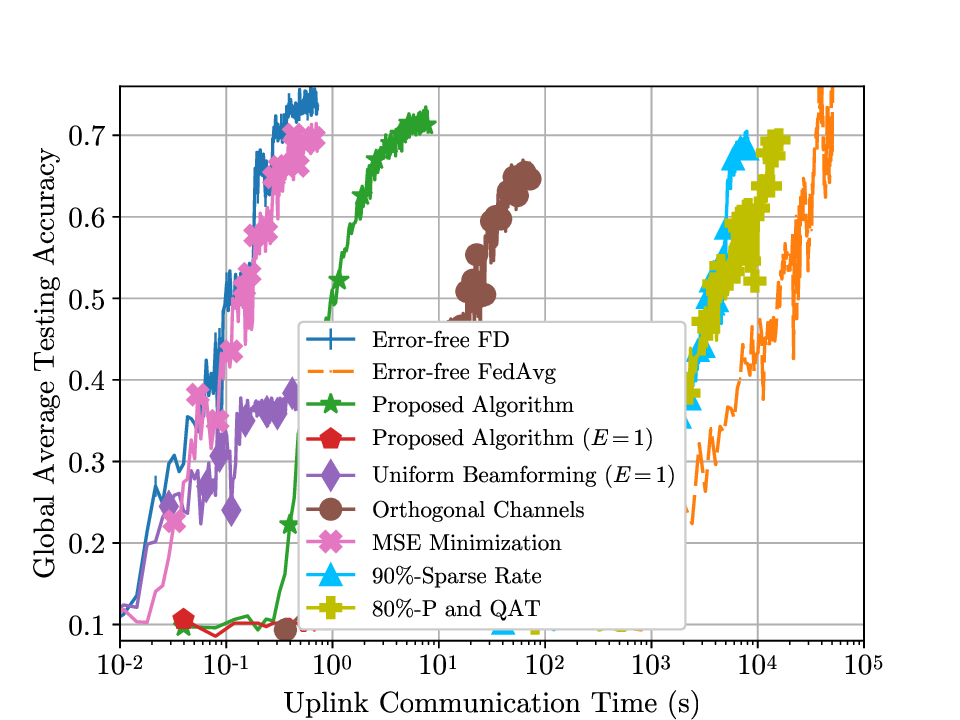}
        \end{minipage}}
    \caption{\textcolor{black}{Average global testing accuracy of ResNet-18 over CIFAR-10 in the non-IID setting.}}
\end{figure}
\textcolor{black}{In Fig. 6 (a), (b), we plot the global testing accuracies of all approaches under ResNet-18 over IID CIFAR-10 datasets with respect to overall time consumption, and uplink communication time, respectively. We first see in Fig. 6 (a) that the proposed algorithm with multiple local epochs approaches the error-free FD performance after 40000 seconds and achieves the highest testing accuracy among the other benchmarks under the restricted time consumption (i.e., smaller than 60000 seconds). Then, we see that the uniform beamforming approach achieves a much smaller accuracy that the proposed algorithm under the same number of local epochs, demonstrating the necessity of the optimal transceiver design in over-the-air FD. Next, the proposed algorithm achieves a higher performance than the MSE minimization benchmark, validating the effectiveness of the analytical convergence gap as the objective function for the optimal transceiver design problem. The proposed algorithm also performs better than the baseline with orthogonal channels due to the benefits of over-the-air computation. For the two communication-efficient over-the-air FL baselines, model pruning achieves the lowest testing accuracy because the model is extremely compressed at WDs. Although compressive sensing attains comparative performance with the proposed algorithm, the complexity of the AMP algorithm is excessively large so that the overall time consumption is longer than the other approaches. In Fig. 6 (b), we can observe that with only local-averaged soft predictions communicated,  the proposed algorithm achieves a higher communication efficiency compared to the FL alternatives. In addition, the proposed over-the-air FD with optimal transceiver design approaches the error-free method with the cost of signaling overhead and execution for the proposed Algorithm 1.}\par
\textcolor{black}{Fig. 7 (a), (b) plot the global testing accuracies of all approaches with respect to the overall time consumption, and uplink communication time, respectively. In Fig. 7 (a), we observe the similar trend as that in the IID case, where the proposed algorithm outperforms all other FD and FL benchmarks with significant communication overhead reduction. Nevertheless, we observe that the performance gap between the error-free FedAvg and error-free FD is widen, aligning with the intuition that online co-distillation underperforms in the non-IID setting. Moreover, the performance gap between error-free FD and the proposed algorithm is larger than that under IID dataset setting, indicating that the non-IID dataset setting enlarges the impact of the imperfect wireless channels. Therefore, an optimal transceiver design to combat the channel fading and noises is essential in the non-IID scenario. From Fig. 7 (b), similar observations regarding the communication efficiencies of all approaches can be obtained as those in the IID dataset setting.}
\section{Conclusions and Future Directions}
This paper investigates the transceiver design for over-the-air FD. We derived the convergence rate of over-the-air FD with a diminishing step size in the closed form. Based on this, we formulated a transceiver optimization problem to maximize the convergence rate while meeting the power constraints of the transceivers. With the optimal transmit power of WDs and estimator for over-the-air aggregation derived in closed forms, we proposed an efficient algorithm to find the optimal receiver beamforming via the SDR technique. We proved that there exists no optimality gap between the original and relaxed problems for the beamforming optimization. We then demonstrated that the proposed over-the-air FD approach not only exhibits superior learning performance, but also significantly reduces the communication overhead compared with conventional FL alternatives.\par
\textcolor{black}{Finally, we conclude the paper with possible future directions. First, it would be interesting to investigate the effectiveness of the proposed optimal transceiver design in over-the-air FD for extremely large models nowadays, e.g., LLMs. In addition, although the local dataset and model parameters are kept at the device side, the disclosure of model soft predictions may also lead to privacy leakage. It is also interesting to investigate the privacy-learning trade-off in differentially private over-the-air FD with a much reduced transmit signal dimension. Moreover, the computational complexity of the proposed optimal transceiver design algorithm scales with the number of WDs and the size of the antenna array. One possible way to improve the scalability for massive-scale deployments is to explore the distributed or learning-based optimization approaches, which can be studied in the future work.}
{\appendices
\section{Proof of Theorem 1}
\begin{proof}
We first bound the expected difference in the empirical local loss function of WD \(i\) between iterations \(t\) and \(t+1\), i.e.,
\begin{equation}
        \mathbb{E}\left[F_i\left(\bm{\theta}_{i,t+1}\right)-F_i\left(\bm{\theta}_{i,t}\right)\right]=B=B_1+B_2,
\end{equation}
where
\begin{equation}
    \begin{aligned}
        B_1&=\mathbb{E}\left[F_i\left(\bm{\theta}_{i,t+1};\left\{\mathbf{q}_{t+1}^k\right\}\right)-F_i\left(\bm{\theta}_{i,t+1};\left\{\mathbf{q}_t^k\right\}\right)\right],\\
        B_2&=\mathbb{E}\left[F_i\left(\bm{\theta}_{i,t+1};\left\{\mathbf{q}_t^k\right\}\right)-F_i\left(\bm{\theta}_{i,t};\left\{\mathbf{q}_t^k\right\}\right)\right].
    \end{aligned}
\end{equation}
By Assumption 1 and Eqn. (4), we have
\begin{equation}
\begin{aligned}
    B_2&\leq\underbrace{-\eta_t\nabla F_i\left(\bm{\theta}_{i,t}\right)^\top\mathbb{E}\left[\nabla F_i\left(\bm{\theta}_{i,t};\left\{\hat{\mathbf{r}}_t^k\right\}\right)\right]}_{C_1}\\
    &\quad+\frac{L_1\eta_t^2}{2}\underbrace{\mathbb{E}\left[\left\|\nabla F_i\left(\bm{\theta}_{i,t};\left\{\hat{\mathbf{r}}_t^k\right\}\right)\right\|^2_2\right]}_{C_2}.
\end{aligned}
\end{equation}
According to (1)-(3),
\begin{equation}
\begin{aligned}
    B_1&=\frac{\gamma}{B_i}\sum_{b=1}^{B_i}\left(\left\|G_{\bm{\theta}_{i,t+1}}\left(\mathbf{u}_{i,b}\right)-\mathbf{q}^{v_{i,b}}_{t+1}\right\|^2_2-\left\|G_{\bm{\theta}_{i,t+1}}\left(\mathbf{u}_{i,b}\right)-\right.\right.\\
    &\qquad\qquad\quad\left.\left.\mathbf{q}^{v_{i,b}}_t\right\|^2_2\right)\\
    &= \frac{\gamma}{B_i}\sum_{b=1}^{B_i}\left(2G_{\bm{\theta}_{i,t+1}}\left(\mathbf{u}_{i,b}\right)^\top\mathbf{q}^{v_{i,b}}_t-2G_{\bm{\theta}_{i,t+1}}\left(\mathbf{u}_{i,b}\right)^\top\mathbf{q}^{v_{i,b}}_{t+1}\right.\\
    &\qquad\qquad\quad\left.+\left\|\mathbf{q}^{v_{i,b}}_{t+1}\right\|^2_2-\left\|\mathbf{q}^{v_{i,b}}_t\right\|^2_2\right)\\
    &=\frac{\gamma}{B_i}\sum_{b=1}^{B_i}\left(\mathbf{q}^{v_{i,b}}_{t+1}+\mathbf{q}^{v_{i,b}}_t-2G_{\bm{\theta}_{i,t+1}}\left(\mathbf{u}_{i,b}\right)\right)^\top\left(\mathbf{q}^{v_{i,b}}_{t+1}-\mathbf{q}^{v_{i,b}}_t\right)\\
    &\stackrel{(a)}{\leq}\frac{\gamma}{B_i}\sum_{b=1}^{B_i}\left\|\mathbf{q}^{v_{i,b}}_{t+1}+\mathbf{q}^{v_{i,b}}_t-2G_{\bm{\theta}_{i,t+1}}\left(\mathbf{u}_{i,b}\right)\right\|_2\left\|\mathbf{q}^{v_{i,b}}_{t+1}-\mathbf{q}^{v_{i,b}}_t\right\|_2\\
    &\stackrel{(b)}{\leq}\frac{\gamma}{B_i}\sum_{b=1}^{B_i}\left(\left\|\mathbf{q}^{v_{i,b}}_{t+1}\right\|_2+\left\|\mathbf{q}^{v_{i,b}}_t\right\|_2+\left\|2G_{\bm{\theta}_{i,t+1}}\left(\mathbf{u}_{i,b}\right)\right\|_2\right)\\
    &\qquad\qquad\quad\times\left\|\mathbf{q}^{v_{i,b}}_{t+1}-\mathbf{q}^{v_{i,b}}_t\right\|_2\\
    &\stackrel{(c)}{\leq}\frac{\gamma}{B_i}\sum_{k=1}^{K}B_i^k4\left\|\mathbf{q}^k_{t+1}-\mathbf{q}^k_t\right\|_2\\
    &=\frac{4\gamma}{B_i}\sum_{k=1}^{K}B_i^k\left\|\frac{1}{B^k}\sum_{i=1}^M\sum_{b\in\mathcal{B}_i^k}\left(G_{\bm{\theta}_{i,t+1}}\left(\mathbf{u}_{i,b}\right)-G_{\bm{\theta}_{i,t}}\left(\mathbf{u}_{i,b}\right)\right)\right\|_2\\
    &\stackrel{(d)}{\leq}\frac{4\gamma}{B_i}\sum_{k=1}^{K}\frac{B_i^k}{B^k}\sum_{i=1}^M\sum_{b\in\mathcal{B}_i^k}L_2\left\|\bm{\theta}_{i,t+1}-\bm{\theta}_{i,t}\right\|_2\\
    &=\frac{4\gamma}{B_i}\sum_{k=1}^{K}\frac{B_i^k}{B^k}\sum_{i=1}^M\sum_{b\in\mathcal{B}_i^k}L_2\left\|-\eta_t\nabla F_i\left(\bm{\theta}_{i,t};\left\{\hat{\mathbf{r}}_t^k\right\}\right)\right\|_2\\
    &\stackrel{(e)}{\leq}4\gamma L_2\eta_tS,
\end{aligned}
\end{equation}
where (\(a\)) is due to Cauchy–Schwarz Inequality. Inequality (\(b\)) follows from the triangle inequality. For the inequality (\(c\)), we first see that the \(l_1\) norms of \(\mathbf{q}^{v_{i,b}}_{t+1}\), \(\mathbf{q}^{v_{i,b}}_t\) and \(G_{\bm{\theta}_{i,t+1}}\left(\mathbf{u}_{i,b}\right)\) are equal to 1 since they are the average of probability vectors (outputs of the softmax layer) whose entries sum to 1. Then according to the property of \(l_1\), \(l_2\) norms, for any vector \(\mathbf{x}\in\mathbb{R}^d\), we have \(\left\|\mathbf{x}\right\|_2\leq\left\|\mathbf{x}\right\|_1\), and the \(l_2\) norms of \(\mathbf{q}^{v_{i,b}}_{t+1}\), \(\mathbf{q}^{v_{i,b}}_t\) and \(G_{\bm{\theta}_{i,t+1}}\left(\mathbf{u}_{i,b}\right)\) are bounded by 1. Subsequently, we interchange the summation of sample indices \(b\) with class indices \(k\). Inequality (\(d\)) is due to Jensen's Inequality and Assumption 2. Inequality (\(e\)) follows from Assumption 3.\par
Next, we bound \(C_1\) and \(C_2\). By (1) and chain rule, we have\\
\begin{equation}
\begin{aligned}
    \nabla F_i\left(\bm{\theta}_{i,t};\left\{\hat{\mathbf{r}}_t^k\right\}\right)&=\nabla F_i\left(\bm{\theta}_{i,t}\right)+\frac{2\gamma}{B_i}\sum_{b=1}^{B_i}\frac{\partial G_{\bm{\theta}_{i,t}}\left(\mathbf{u}_{i,b}\right)}{\partial\bm{\theta}_{i,t}}\\
    &\quad\times\left(G_{\bm{\theta}_{i,t}}\left(\mathbf{u}_{i,b}\right)-\hat{\mathbf{r}}_t^{v_{i,b}}\right)\\
    &=\nabla F_i\left(\bm{\theta}_{i,t}\right)+\frac{2\gamma}{B_i}\sum_{b=1}^{B_i}\frac{\partial G_{\bm{\theta}_{i,t}}\left(\mathbf{u}_{i,b}\right)}{\partial\bm{\theta}_{i,t}}\\
    &\quad\times\left(G_{\bm{\theta}_{i,t}}\left(\mathbf{u}_{i,b}\right)-\hat{\mathbf{r}}_t^{v_{i,b}}+\mathbf{q}_t^{v_{i,b}}-\mathbf{q}_t^{v_{i,b}}\right)\\
    &=\nabla F_i\left(\bm{\theta}_{i,t}\right)+\frac{2\gamma}{B_i}\sum_{b=1}^{B_i}\frac{\partial G_{\bm{\theta}_{i,t}}\left(\mathbf{u}_{i,b}\right)}{\partial\bm{\theta}_{i,t}}\\
    &\quad\times\left(\mathbf{q}_t^{v_{i,b}}-\hat{\mathbf{r}}_t^{v_{i,b}}\right).
\end{aligned}
\end{equation}
By plugging (31) into \(C_1\), we get
\begin{equation}
    \begin{aligned}
        C_1&=-\eta_t\left\|\nabla F_i\left(\bm{\theta}_{i,t}\right)\right\|^2_2-\frac{2\eta_t\gamma}{B_i}\sum_{b=1}^{B_i}\nabla F_i\left(\bm{\theta}_{i,t}\right)^\top\frac{\partial G_{\bm{\theta}_{i,t}}\left(\mathbf{u}_{i,b}\right)}{\partial\bm{\theta}_{i,t}}\\
        &\quad\times\left(\mathbf{q}_t^{v_{i,b}}-\mathbb{E}\left[\hat{\mathbf{r}}_t^{v_{i,b}}\right]\right)\\
        &\stackrel{(f)}{\leq}-\eta_t\left\|\nabla F_i\left(\bm{\theta}_{i,t}\right)\right\|^2_2+2\gamma L_2\eta_t\left\|\nabla F_i\left(\bm{\theta}_{i,t}\right)\right\|_2\sum_{k=1}^K\frac{B_i^k}{B_i}\\
        &\quad\times\left\|\mathbf{q}_t^k-\mathbb{E}\left[\hat{\mathbf{r}}_t^k\right]\right\|_2\\
        &=-\eta_t\left\|\nabla F_i\left(\bm{\theta}_{i,t}\right)\right\|^2_2+2\gamma L_2\eta_t\left\|\nabla F_i\left(\bm{\theta}_{i,t}\right)\right\|_2\Phi_{1,i,t},
    \end{aligned}
\end{equation}
where (\(f\)) is due to the Cauchy–Schwarz Inequality that
\begin{equation}
\begin{aligned}
    &\nabla F_i\left(\bm{\theta}_{i,t}\right)^\top\frac{\partial G_{\bm{\theta}_{i,t}}\left(\mathbf{u}_{i,b}\right)}{\partial\bm{\theta}_{i,t}}\left(\mathbf{q}_t^{v_{i,b}}-\mathbb{E}\left[\hat{\mathbf{r}}_t^{v_{i,b}}\right]\right)\geq\\
    &-\left\|\nabla F_i\left(\bm{\theta}_{i,t}\right)\right\|_2\left\|\frac{\partial G_{\bm{\theta}_{i,t}}\left(\mathbf{u}_{i,b}\right)}{\partial\bm{\theta}_{i,t}}\right\|_2\left\|\mathbf{q}_t^{v_{i,b}}-\mathbb{E}\left[\hat{\mathbf{r}}_t^{v_{i,b}}\right]\right\|_2.
\end{aligned}
\end{equation}
The inequality is also due to the property of the matrix \(2-\)norm that for any matrix \(\mathbf{A}\in\mathbb{R}^{M\times N}\) and vector \(\mathbf{x}\in\mathbb{R}^N\), \(\left\|\mathbf{A}\mathbf{x}\right\|_2\leq\left\|\mathbf{A}\right\|_2\left\|\mathbf{x}\right\|_2\). Notice that by Assumption 2, we can bound the norm of the partial derivative of the learned model function \(G_{\bm{\theta}_{i,t}}\left(\cdot\right)\) by the Lipschitz constant \(L_2\). The last equality follows from the zero mean of \(\hat{\mathbf{n}}_t^k\) and the definition of \(\Phi_{1,i,t}\) in (18).\par
Similarly, with Eqn. (31), we have
\begin{equation}
    \begin{aligned}
        C_2 &= \left\|\nabla F_i\left(\bm{\theta}_{i,t}\right)\right\|^2_2+\frac{4\gamma}{B_i}\sum_{b=1}^{B_i}\nabla F_i\left(\bm{\theta}_{i,t}\right)^\top\frac{\partial G_{\bm{\theta}_{i,t}}\left(\mathbf{u}_{i,b}\right)}{\partial\bm{\theta}_{i,t}}\left(\mathbf{q}_t^{v_{i,b}}-\right.\\
        &\quad\left.\mathbb{E}\left[\hat{\mathbf{r}}_t^{v_{i,b}}\right]\right)+\mathbb{E}\left[\left\|\frac{2\gamma}{B_i}\sum_{b=1}^{B_i}\frac{\partial G_{\bm{\theta}_{i,t}}\left(\mathbf{u}_{i,b}\right)}{\partial\bm{\theta}_{i,t}}\left(\mathbf{q}_t^{v_{i,b}}-\hat{\mathbf{r}}_t^{v_{i,b}}\right)\right\|^2_2\right]\nonumber
                        \end{aligned}
\end{equation}
\begin{equation}
\begin{aligned}
                &\stackrel{(g)}{\leq}\left\|\nabla F_i\left(\bm{\theta}_{i,t}\right)\right\|^2_2+4\gamma L_2\left\|\nabla F_i\left(\bm{\theta}_{i,t}\right)\right\|_2\Phi_{1,i,t}\\
        &\quad+\mathbb{E}\left[\left\|\frac{2\gamma}{B_i}\sum_{b=1}^{B_i}\frac{\partial G_{\bm{\theta}_{i,t}}\left(\mathbf{u}_{i,b}\right)}{\partial\bm{\theta}_{i,t}}\left(\mathbf{q}_t^{v_{i,b}}-\hat{\mathbf{r}}_t^{v_{i,b}}\right)\right\|^2_2\right]\\
        &\stackrel{(h)}{\leq}\left\|\nabla F_i\left(\bm{\theta}_{i,t}\right)\right\|^2_2+4\gamma L_2\left\|\nabla F_i\left(\bm{\theta}_{i,t}\right)\right\|_2\Phi_{1,i,t}\\
        &\quad+4\gamma^2L_2^2\left(\Phi_{1,i,t}^2+\Phi_{2,i,t}^2\right),
            \end{aligned}
        \end{equation}
where (\(g\)) follows from (32). The inequality (\(h\)) is due to Jensen's Inequality as well as Assumption 2. By the independence and the zero mean of \(\hat{\mathbf{n}}_t^k\), \(\mathbb{E}\left[\sum_{k=1}^{K}\frac{B_i^k}{B_i}\left\|\mathbf{q}_t^{v_{i,b}}-\hat{\mathbf{r}}_t^{v_{i,b}}\right\|^2_2\right]\) can be decomposed into \(\Phi_{1,i,t}^2+\Phi_{2,i,t}^2\).\par
Plug (29), (30), (32) and (34) back to (27), we get
\begin{equation}
    \begin{aligned}
        B&\leq \left(\frac{\eta_t^2L_1}{2}-\eta_t\right)\left\|\nabla F_i\left(\bm{\theta}_{i,t}\right)\right\|^2_2+L_1L_2^2\eta_t^2\left(\Phi_{1,i,t}^2+\Phi_{2,i,t}^2\right)\\
        &\quad\times2\gamma^2+2\gamma L_2\eta_t\left(L_1\eta_t+1\right)\left\|\nabla F_i\left(\bm{\theta}_{i,t}\right)\right\|_2\Phi_{1,i,t}\\
        &\quad+4\gamma L_2\eta_tS\\
        &\leq -\frac{\eta_t}{2}\left\|\nabla F_i\left(\bm{\theta}_{i,t}\right)\right\|^2_2+2\gamma L_2\eta_t\left(L_1\eta_t+1\right)\left\|\nabla F_i\left(\bm{\theta}_{i,t}\right)\right\|_2\\
    &\quad\times\Phi_{1,i,t}+2\gamma^2L_1L_2^2\eta_t^2\left(\Phi_{1,i,t}^2+\Phi_{2,i,t}^2\right)+4\gamma L_2\eta_tS.
    \end{aligned}
\end{equation}
The last inequality is due to \(-\eta_t+\frac{\eta_t^2L_1}{2}\leq -\frac{\eta_t}{2}\) with \(\eta_t\leq \frac{1}{L_1}\). We rearrange the terms, divide both sides by \(\frac{\eta_t^2}{2}\) and sum over \(t=0\) to \(T-1\) to obtain
\begin{equation}
    \begin{aligned}
        \underbrace{\sum_{t=0}^{T-1}\frac{1}{\eta_t}\left\|\nabla F_i\left(\bm{\theta}_{i,t}\right)\right\|^2_2}_{C_3}&\leq \sum_{t=0}^{T-1}\frac{2}{\eta_t^2}\mathbb{E}\left[F_i\left(\bm{\theta}_{i,t}\right)-F_i\left(\bm{\theta}_{i,t+1}\right)\right]\\
        &\quad+\sum_{t=0}^{T-1}\frac{4}{\eta_t}\gamma L_2\left(L_1\eta_t+1\right)\left\|\nabla F_i\left(\bm{\theta}_{i,t}\right)\right\|_2\\
        &\quad\times\Phi_{1,i,t}+\sum_{t=0}^{T-1}\frac{8}{\eta_t}\gamma L_2S\\
        &\quad+\sum_{t=0}^{T-1}4\gamma^2L_1L_2^2\left(\Phi_{1,i,t}^2+\Phi_{2,i,t}^2\right).
        \end{aligned}
\end{equation}
According to the proof of Theorem 3.5 in \cite{ref12},
\begin{equation}
    \begin{aligned}
        C_3&\leq\frac{2}{\eta_0^2}\left[F_i\left(\bm{\theta}_{i,0}\right)+\sum_{t=1}^{T-1}\left(t-\left(t-1\right)\right)\mathbb{E}\left[F_i\left(\bm{\theta}_{i,t}\right)\right]\right]\\
        &\quad+\sum_{t=0}^{T-1}\frac{8}{\eta_t}\gamma L_2S+\sum_{t=0}^{T-1}4\gamma^2L_1L_2^2\left(\Phi_{1,i,t}^2+\Phi_{2,i,t}^2\right)\\
        &\quad+\sum_{t=0}^{T-1}\frac{4}{\eta_t}\gamma L_2\left(L_1\eta_t+1\right)\left\|\nabla F_i\left(\bm{\theta}_{i,t}\right)\right\|_2\Phi_{1,i,t}\\
        &\leq\frac{2Tf_{i,max}}{\eta_0^2}+\sum_{t=0}^{T-1}\frac{4}{\eta_t}\gamma L_2\left(L_1\eta_t+1\right)\left\|\nabla F_i\left(\bm{\theta}_{i,t}\right)\right\|_2\Phi_{1,i,t}\\
        &\quad+\sum_{t=0}^{T-1}\frac{8}{\eta_t}\gamma L_2S+\sum_{t=0}^{T-1}4\gamma^2L_1L_2^2\left(\Phi_{1,i,t}^2+\Phi_{2,i,t}^2\right).
    \end{aligned}
\end{equation}
Recall that \(\hat{\bm{\theta}}_{i,T}\) is randomly chosen from \(\left\{\bm{\theta}_{i,t},\forall t\right\}\) at all the previous iterations with probability \(P\left(\hat{\bm{\theta}}_{i,T}=\bm{\theta}_{i,t}\right)=\frac{1/\eta_t}{\sum_{t=0}^{T-1}1/\eta_t}\). We divide both sides by \(\sum_{t=0}^{T-1}\frac{1}{\eta_t}\), which gives
\begin{equation}
\begin{aligned}
        \mathbb{E}\left[\left\|\nabla F_i\left(\hat{\bm{\theta}}_{i,T}\right)\right\|^2_2\right]&\leq\frac{3f_{i,max}}{\eta_0\sqrt{T}}+\sum_{t=0}^{T-1}\frac{6\gamma\eta_0L_2\left(L_1\eta_t+1\right)}{\eta_t}\\
                &\quad\times\frac{\left\|\nabla F_i(\bm{\theta}_{i,t})\right\|_2\Phi_{1,i,t}}{T^{\frac{3}{2}}}+8\gamma L_2S\\
        &\quad+\sum_{t=0}^{T-1}6\eta_0\gamma^2L_2^2L_1\left(\frac{\Phi_{1,i,t}^2+\Phi_{2,i,t}^2}{T^{\frac{3}{2}}}\right)
            \end{aligned}
        \end{equation}
where the inequality holds because \(\sum_{t=0}^{T-1}\frac{1}{\eta_t}\geq \frac{1}{\eta_0}\int_{t=0}^T\sqrt{t}\,dt=\frac{2}{3\eta_0}T^{\frac{3}{2}}\). The proof of Theorem 1 is thus completed.
\end{proof}
\section{Proof of Proposition 1}
\begin{proof}
For each training round \(t\), we see that
\begin{equation}
\begin{aligned}
    &\sum_{i=1}^M\sum_{t=0}^{T-1}A_1\frac{\left(L_1\eta_t+1\right)\left\|\nabla F_i(\bm{\theta}_{i,t})\right\|_2\Phi_{1,i,t}}{\eta_tT^{\frac{3}{2}}}+\sum_{i=1}^M\sum_{t=0}^{T-1}A_2\\
    &\times\frac{\Phi_{1,i,t}^2+\Phi_{2,i,t}^2}{T^{\frac{3}{2}}}\stackrel{(a)}{\geq}\sum_{i=1}^M\sum_{t=0}^{T-1}A_2\frac{\Phi_{2,i,t}^2}{T^{\frac{3}{2}}}.
\end{aligned}
\end{equation}
The equality \((a)\) holds if \(\Phi_{1,i,t}=0,\forall i,t\), which gives the following two equations:
\begin{equation}
    \begin{aligned}
        \frac{\mathbf{w}_t^H\mathbf{h}_{i,t}P_{i,t}^{k^\ast}}{\lambda_t^k\hat{q}_{i,t}^k}-\frac{B_i^k}{B^k}&=0,\quad\forall i,t,k,\\
        a_{i,t}^{k^\ast}-\frac{\mathbf{w}_t^H\mathbf{h}_{i,t}P_{i,t}^{k^\ast}}{\lambda_t^k\hat{q}_{i,t}^k}&=0,\quad\forall i,t,k.
    \end{aligned}
\end{equation}
The optimal solution pair can be solved as
\begin{equation}
    \begin{aligned}
        P_{i,t}^{k^\ast} &= \frac{B^k_i\lambda_t^k\hat{q}_{i,t}^k\left(\mathbf{w}_t^H\mathbf{h}_{i,t}\right)^H}{B^k\left|\mathbf{w}_t^H\mathbf{h}_{i,t}\right|^2},\quad\forall i,t,k,\\
        a_{i,t}^{k^\ast} &= \frac{B_i^k}{B^k},\quad\forall i,t,k.
    \end{aligned}
\end{equation}
Then, by (41) and the peak transmit power constraint in (9), we have
\begin{equation}
\lambda_t^{k^2}=\frac{B^{k^2}\left|\mathbf{w}_t^H\mathbf{h}_{i,t}\right|^2\left|P_{i,t}^k\right|^2}{B_i^{k^2}\hat{q}_{i,t}^{k^2}}\leq\frac{B^{k^2}\left|\mathbf{w}_t^H\mathbf{h}_{i,t}\right|^2P_i}{B_i^{k^2}\hat{q}_{i,t}^{k^2}}, \quad\forall i,t,k.
\end{equation}
Given the receiver beamforming vector \(\mathbf{w}_t\), by plugging in the optimal solutions of \(\left\{P_{i,t}^k,a^k_{i,t},\forall i,t,k\right\}\), Problem (P1) can be transformed into the following problem.
\begin{equation}
    \begin{aligned}
        \mbox{(P5)}\quad& \underset{\left\{\lambda_t^k,\forall t,k\right\}}{\text{min}}
& & \sum_{i=1}^M\sum_{t=0}^{T-1}A_2\frac{\Phi_{2,i,t}^2}{T^{\frac{3}{2}}}\\
 & \text{s.t.} & & \lambda_t^k\leq\frac{B^k\left|\mathbf{w}_t^H\mathbf{h}_{i,t}\right|\sqrt{P_i}}{B^k_i\hat{q}_{i,t}^k},\quad \forall i,k,t.
    \end{aligned}
\end{equation}
By analyzing the objective function of (43), we see that
\begin{equation}
    \begin{aligned}
        \sum_{i=1}^M\sum_{t=0}^{T-1}A_2\frac{\Phi_{2,i,t}^2}{T^{\frac{3}{2}}}&=\sum_{i=1}^M\sum_{t=0}^{T-1}\frac{A_2\sum_{k=1}^K\frac{B^k_i}{B_i}\mathbb{E}\left[\left\|\hat{\mathbf{n}}_t^k\right\|^2_2\right]}{{\lambda_t^k}^2T^{\frac{3}{2}}}\\
        &=\sum_{i=1}^M\sum_{t=0}^{T-1}\frac{A_2\sum_{k=1}^K\frac{B^k_i}{B_i}K\sigma_n^2}{{\lambda_t^k}^2T^{\frac{3}{2}}}\\
        &=\sum_{i=1}^M\frac{A_2K\sigma_n^2}{{\lambda_t^k}^2\sqrt{T}}.
    \end{aligned}
\end{equation}
The second equality follows from \(\left\|\mathbf{w}_t\right\|^2_2=1,\forall t\) and \(\mathbf{n}_t[d]\sim\mathcal{CN}\left(\mathbf{0},\sigma_n^2\cdot \mathbf{I}\right)\). Therefore, the optimal \(\lambda_t^k\) is given by
\begin{equation}
\lambda_t^{k^\ast}=\min_{i\in\mathcal{M}}\frac{B^k\left|\mathbf{w}_t^H\mathbf{h}_{i,t}\right|\sqrt{P_i}}{B^k_i\hat{q}_{i,t}^k},\quad \forall k,t.
\end{equation}
The proof of Proposition 1 is thus completed.
\end{proof}
\section{Proof of Proposition 2}
\begin{proof}
With Proposition 1, we are ready to solve the optimal receiver beamforming design problem per round in Problem (P4). By relaxing the rank-one constraint, the semidefinite optimization problem is
\begin{subequations}
    \begin{align}
        \mbox{(P6)}\quad& \underset{\left\{\mathbf{W}_t\in\mathcal{H}^N,\left\{e^k\right\}_{k=1}^K\right\}}{\text{min}}
& & \Omega\left(\mathbf{W}_t\right)=\sum_{k=1}^K\frac{e^k}{B^k}\sum_{i=1}^M\frac{B_i^k}{B_i}\tag{46}\\
 & \text{s.t.} & & \mbox{Tr}\left(\mathbf{W}_t\right)=1,\tag{46a}\\
 &&&\mathbf{W}_t\succeq\mathbf{0},\tag{46b}\\
 &&&e^1+\mbox{Tr}\left(\mathbf{W}_t\mathbf{H}_{j,t}^1\right)\geq 0,\quad\forall j,\tag{46c.1}\\
 &&&\cdots\notag\\
 &&& e^K+\mbox{Tr}\left(\mathbf{W}_t\mathbf{H}_{j,t}^K\right)\geq 0,\quad\forall j.\tag{46c.\(K\)}
    \end{align}
\end{subequations}\par
First of all, we have the following two lemmas on Problem (P6) to facilitate the proof of Proposition 2.\par
\textit{\textbf{Lemma C.1:}} There always exists an optimal solution to Problem (P6).
\begin{proof}
    We first see that the constraints (46a), (46c.1), ..., (46c.\(K\)) are affine and their respective feasible sets are non-empty and closed. In addition, the set of positive semidefinite hermitian matrices associated with the constraint (46b) is convex, non-empty and closed. Accordingly, given that the intersection of closed sets is closed, the feasible set of Problem (P6) is non-empty and closed. Moreover, the affine objective function (46) is continuous and thus coercive since it approaches infinity when the set \(\left\{e^k\right\}_{k=1}^K\) approaches infinity. Therefore, according to Weierstrass Extreme Value Theorem, there always exists an optimal solution to Problem (P6).
\end{proof}
\par
\textit{\textbf{Lemma C.2:}} Within each constraint (46c.\(k\)), for any optimal solution \(\left\{\mathbf{W}_t^\ast,\left\{e^{k^\ast}\right\}_{k=1}^K\right\}\) to Problem (P6), at least one constraint achieves equality.
\begin{proof}
    We show this by contradiction. Assume that there exists an optimal solution \(\left\{\left\{e^{k^\ast}\right\}_{k=1}^K,\mathbf{W}_t^\ast\right\}\) with one constraint (46c.\(\hat{k}\)) satisfying \(e^{\hat{k}^\ast}+\mbox{Tr}\left(\mathbf{W}_t^\ast\mathbf{H}_{j,t}^{\hat{k}}\right)> 0,\forall j\). Then, we can always find a \(\hat{e}^{\hat{k}}<e^{\hat{k}^\ast}\) with \(\hat{e}^{\hat{k}}+\max_{j\in\mathcal{M}}\mbox{Tr}\left(\mathbf{W}_t^\ast\mathbf{H}_{j,t}^{\hat{k}}\right)= 0\). As a result, the objective value achieved by the solution \(\left\{\left\{e^{k^\ast}\right\}_{k\neq\hat{k}},\hat{e}^{\hat{k}},\mathbf{W}_t^\ast\right\}\) is strictly lower than that by \(\left\{\left\{e^{k^\ast}\right\}_{k=1}^K,\mathbf{W}_t^\ast\right\}\). This contradicts with the assumption. Therefore, at least one constraint within each (46c.\(k\)) achieves equality with any optimal solution to Problem (P6).
\end{proof}
\par
According to Lemmas C.1 and C.2, we prove Proposition 2 by contradiction. Assume that there exists an optimal solution \(\mathbf{W}^\ast_t\) to (P6) with rank \(R>1\). We decompose the positive semidefinite matrix \(\mathbf{W}^\ast_t\) by \(\mathbf{W}^\ast_t=\mathbf{V}\mathbf{\Lambda}\mathbf{V}^H\), where \(\mathbf{V}\in\mathbb{C}^{N\times N}\) is a unitary matrix with columns \(\mathbf{v}_1,\mathbf{v}_2,\cdots,\mathbf{v}_N\). \(\mathbf{\Lambda}\in\mathbb{R}^{N\times N}\) is a diagonal matrix with non-zero leading diagonal entries \(\Lambda_1,\Lambda_2,\cdots,\Lambda_R\), which are eigenvalues of \(\mathbf{W}^\ast_t\). We have \(\sum_{r=1}^R\Lambda_r=1\) according to the constraint (46a). Moreover, since the columns of \(\mathbf{V}\) span the complex space \(\mathbb{C}^N\), for each device \(i\), its channel coefficient \(\mathbf{h}_{i,t}\) can be represented as a linear combination of columns of \(\mathbf{V}\), i.e.,
\begin{equation}
    \mathbf{h}_{i,t}=\sum_{n=1}^N z_{i,n}\mathbf{v}_n,
\end{equation}
where \(z_{i,n}=\mathbf{h}_{i,t}^H\mathbf{v}_n\in\mathbb{C}\) is the coefficient of the \(n\)-th component. Without loss of generality, we assume that the first constraint within each constraint (46c.\(k\)) in (P6) achieves equality under the optimal solution according to Lemma C.2,
\begin{equation}
\begin{aligned}
    &e^{k^\ast}+\frac{P_1}{\left(B^k_1\hat{q}_{1,t}^k\right)^2}\mathbf{h}_{1,t}^H\mathbf{V}\mathbf{\Lambda}\mathbf{V}^H\mathbf{h}_{1,t}= 0,\quad\forall k\\
    &e^{k^\ast}+\frac{P_i}{\left(B^k_i\hat{q}_{i,t}^k\right)^2}\mathbf{h}_{i,t}^H\mathbf{V}\mathbf{\Lambda}\mathbf{V}^H\mathbf{h}_{i,t}\geq 0, \quad\forall i\neq 1,k.\\
\end{aligned}
\end{equation}
Based on Eqn. (48), we solve the optimal \(\left\{e^{k^\ast},\forall k\right\}\) as
\begin{equation}
    e^{k^\ast}=-\frac{P_1}{\left(B^k_1\hat{q}_{1,t}^k\right)^2}\sum_{r=1}^R \left|z_{1,r}\right|^2\Lambda_r,\quad\forall k.
\end{equation}
Accordingly, the optimal objective value is
\begin{equation}
    \Omega\left(\mathbf{W}_t^\ast\right)=-\sum_{k=1}^K\sum_{i=1}^M\frac{B_i^kP_1}{B_iB^k\left(B^k_1\hat{q}_{1,t}^k\right)^2}\sum_{r=1}^R \left|z_{1,r}\right|^2\Lambda_r.
\end{equation}\par
Then, we construct a feasible solution \(\hat{\mathbf{W}}_t=\hat{\mathbf{V}}\hat{\mathbf{\Lambda}}\hat{\mathbf{V}}^H\), where \(\hat{\mathbf{\Lambda}}\in\mathbb{R}^{N\times N}\) is a diagonal matrix with only one non-zero leading diagonal entry, which is \(\hat{\Lambda}_1 = 1\). The first column of \(\hat{\mathbf{V}}\) is constructed as \(\hat{\mathbf{v}}=\sum_{r=1}^R\sqrt{\Lambda_r}\mathbf{v}_r\). We can verify that \(\hat{\mathbf{v}}\) is a unit vector by \(\left\|\hat{\mathbf{v}}\right\|_2^2=\sum_{r=1}^R\Lambda_r=1\). According to Gram–Schmidt process, we can always find the other \(N-1\) vectors so that they can formulate an orthonormal basis together with \(\hat{\mathbf{v}}\). Without loss of generality, we assume that the first element of \(\hat{\mathbf{v}}\) is non-zero. Then, one simple approach is to leverage unit vectors \(\mathbf{u}_2=\left[0,1,0,0,\cdots,0\right]^\top\), \(\mathbf{u}_3=\left[0,0,1,\cdots,0\right]^\top\), \(\cdots\), \(\mathbf{u}_K=\left[0,0,0,\cdots,1\right]^\top\), together with \(\hat{\mathbf{v}}\), for the construction of matrix \(\hat{\mathbf{V}}\), i.e.,
\begin{equation}
    \hat{\mathbf{V}}=\left[\hat{\mathbf{v}},\mathbf{u}_2,\cdots,\mathbf{u}_K\right].
\end{equation}
Notice that the constructed \(\hat{\mathbf{W}}_t\) is feasible to Problem (P6) since \(\mbox{Tr}\left(\hat{\mathbf{W}}_t\right)=1\) and \(\hat{\mathbf{W}}_t\) is positive semidefinite. Moreover, by Cauchy–Schwarz Inequality, we have
\begin{equation}
    \left|\sum_{r=1}^Rz_{i,r}\sqrt{\Lambda_r}\right|^2\geq \sum_{r=1}^R \left|z_{i,r}\right|^2\Lambda_r, \quad \forall i.
\end{equation}
According to (48) and (52),
\begin{equation}
\begin{aligned}
    e^{k^\ast}+\frac{P_i}{\left(B^k_i\hat{q}_{i,t}^k\right)^2}\sum_{r=1}^R \left|z_{i,r}\right|^2\Lambda_r&\geq0,\quad\forall i,k,\\
    e^{k^\ast}+\mbox{Tr}\left(\hat{\mathbf{W}}_t\mathbf{H}_{i,t}^k\right)&\geq0,\quad\forall i,k.
\end{aligned}
\end{equation}
In this way, the constructed \(\hat{\mathbf{W}}_t\) is feasible for Problem (P6). Then, we calculate the objective value associated with the constructed \(\hat{\mathbf{W}_t}\), i.e.,
\begin{equation}
        \Omega\left(\hat{\mathbf{W}}_t\right)=-\sum_{k=1}^K\sum_{i=1}^M\frac{B_i^kP_1}{B_iB^k\left(B^k_1\hat{q}_{1,t}^k\right)^2}\left|\sum_{r=1}^Rz_{1,r}\sqrt{\Lambda_r}\right|^2.
\end{equation}
According to (50) and (52), we can see that
\begin{equation}
    \Omega\left(\hat{\mathbf{W}}_t\right)\leq\Omega\left(\mathbf{W}_t^\ast\right).
\end{equation}
The equality is achieved if and only if at least \(R-1\) elements in \(\left\{z_{1,r}\sqrt{\Lambda_r}\right\}_{r=1}^R\) are equal to 0. First, if all of the elements are equal to 0, then \(\mathbf{h}_{i,t}=\mathbf{0}\), which is trivial. Notice that for \(\mathbf{W}_t^\ast\) with rank \(R>1\), \(\left\{\Lambda_r\right\}_{r=1}^R\) are all non-zero. In the following, two subcases are discussed.
\begin{itemize}
    \item If at most \(R-2\) elements in \(\left\{z_{1,r}\right\}_{r=1}^R\) are equal to 0, then \(\Omega\left(\hat{\mathbf{W}}_t\right)<\Omega\left(\mathbf{W}_t^\ast\right)\). This contradicts the assumption.
    \item If \(R-1\) elements in \(\left\{z_{1,r}\right\}_{r=1}^R\) are equal to 0, then, based on (47), \(\mathbf{h}_{1,t}\) is linearly dependent with one column of \(\mathbf{V}\). Without loss of generality, we assume that \(\mathbf{h}_{1,t}\) is linearly dependent with \(\mathbf{v}_1\). Subsequently, the optimal objective value becomes
    \begin{equation}
        \Omega\left(\mathbf{W}_t^\ast\right)=-\sum_{k=1}^K\sum_{i=1}^M\frac{B_i^kP_1}{B_iB^k\left(B^k_1\hat{q}_{1,t}^k\right)^2}\left|z_{1,1}\right|^2\Lambda_1.
    \end{equation}
    Similarly, we can construct a feasible solution \(\widetilde{\mathbf{W}}_t=\mathbf{V}\hat{\mathbf{\Lambda}}\mathbf{V}^H\). The objective value associated with \(\widetilde{\mathbf{W}}_t\) is
    \begin{equation}
        \Omega\left(\widetilde{\mathbf{W}}_t\right)=-\sum_{k=1}^K\sum_{i=1}^M\frac{B_i^kP_1}{B_iB^k\left(B^k_1\hat{q}_{1,t}^k\right)^2}\left|z_{1,1}\right|^2.
    \end{equation}
    Since \(\left\{\Lambda_r\right\}_{r=1}^R\) are all non-zero and \(\sum_{r=1}^R\Lambda_r=1\), we have \(\Lambda_1<1\). Therefore, \(\Omega\left(\widetilde{\mathbf{W}}_t\right)<\Omega\left(\mathbf{W}_t^\ast\right)\). This contradicts with the assumption.
\end{itemize}
The proof of Proposition 2 is thus completed.
\end{proof}
}

\end{document}